\documentclass[amsmath,superscriptaddress,aps,pra,twocolumn]{revtex4-2}

\usepackage{graphicx}
\usepackage{dcolumn}
\usepackage{bm}

\usepackage{booktabs}
\usepackage{microtype}
\usepackage{bm}
\usepackage{epsfig}
\usepackage{graphicx}
\usepackage{amssymb,amsmath,amsbsy,amsgen,amsfonts}  
\usepackage{multirow}
\usepackage{dcolumn}
\usepackage{amsthm}
\usepackage{mathrsfs}
\usepackage{latexsym}
\usepackage{array}
\usepackage{color}
\usepackage{amstext}
\allowdisplaybreaks[1]
\usepackage{txfonts}
\usepackage{pifont}
\usepackage{braket}
\usepackage{mathtools}
\usepackage{epstopdf} 
\usepackage[title]{appendix}
\usepackage{color}
\usepackage[dvipsnames]{xcolor}
\usepackage[hidelinks,colorlinks=true, linkcolor=WildStrawberry, urlcolor=WildStrawberry, citecolor=Emerald, linktocpage=true, breaklinks=true, bookmarksopen=true]{hyperref}
\usepackage[USenglish]{babel}

\newcommand{\be}{\begin{equation}}
\newcommand{\ee}{\end{equation}}
\newcommand{\ba}{\begin{array}}
\newcommand{\ea}{\end{array}}
\newcommand{\bqa}{\begin{eqnarray}}
\newcommand{\eqa}{\end{eqnarray}}

\graphicspath{ {Plots/} }
\DeclareSymbolFont{symbols}{OMS}{cmsy}{m}{n}

\begin{document}
\title[]{An effective microscopic model for plasmonic sensing of malaria}

\author{A. S. Kiyumbi}
\email{amos.kiyumbi@udsm.ac.tz}
\affiliation{Department of Physics, Mathematics and Informatics, Dar es Salaam University College of Education, PO Box 2329, Dar es Salaam, Tanzania}
\affiliation{Department of Physics, Stellenbosch University, Private Bag X1, Matieland 7602, South Africa}
\author{M. S. Tame}
\affiliation{Department of Physics, Stellenbosch University, Private Bag X1, Matieland 7602, South Africa}


\begin{abstract}
Malaria remains a major health threat in low-resource regions and rapid diagnostic tests often lack the sensitivity required for early detection. To address this and help establish more sensitive testing devices, we develop a predictive microscopic model for plasmonic biosensing using metasurfaces. Specifically, we consider the detection of plasmodium lactate dehydrogenase (pLDH), a well-known malaria biomarker. An example metasurface is studied to showcase the effective microscopic model -- it consists of a gold nanohole array (150 nm film; 150 nm diameter; 400 nm period) and the biochemistry above it is modelled as stacks of closely packed adlayers. Using Maxwell Garnett effective medium theory we link the refractive index of the pLDH  biomarker adsorbed layer on top of the metasurface to the bulk concentration of pLDH in the buffer. This effective microscopic model accounts for the combined optical properties of the biochemistry matrix, bound pLDH, and the buffer medium. By simulating the sensor using the finite element method and an approximate analytical method, we show that the effective model allows one to determine the sensor response, predict binding interactions, and quantify concentration changes on the sensor surface. We then calculate the sensor sensitivity for our example metasurface and its theoretical limit of detection (LOD) based on spectral and intensity interrogation. The lowest LOD calculated based on the model is $0.02$~nM of pLDH, equivalent to $0.7$~ng/mL, which is a 30 times improvement over current rapid diagnostic tests. While this improvement in performance is highly promising, further work on transferring the ideal theory developed here to field-tested empirical performance will be required. The effective microscopic model we introduce is quite general and the framework developed offers a broadly applicable tool for the design and optimization of other types of highly sensitive plasmonic biosensors.
\end{abstract}

\maketitle

\section{Introduction}
Malaria, a mosquito-borne tropical disease caused by plasmodium parasites has infected humans and caused death for centuries. In a given year malaria results in approximately $500,000$ deaths globally, with more than 95\% occurring within the African continent~\cite{WHO2023, WHO2024, Sibomana2025}. Although new vaccines are being developed and have started to be rolled out in the Ivory Coast as a pilot program, it is expected to take 10 to 20 years to eliminate and completely eradicate malaria~\cite{Sibomana2025}, while the economic burden due to malaria disease--surveillance, monitoring, control, and prevention--in developing countries is increasing~\cite{Andrade2022}. New diagnostic tools and strategies are urgently needed to help fight malaria~\cite{Spackova2016, nguyen2015surface}.

Current conventional diagnosis techniques, such as light microscopy, clinical diagnosis, antigen rapid detection tests, and molecular tests ($\textit{e.g.}$ PCR), have clinical and economic limitations, including precision, affordability, portability, and the necessity of highly trained pathologists~\cite{tangpukdee2009malaria}. On the other hand, optical biosensors using surface plasmon resonance (SPR) have the potential to provide rapid, precise, and label-free detection of malaria protein biomarkers in whole-blood lysate~\cite{Homola2008, Homola2006}. Protein-based planar SPR assays have already found application in malaria diagnostics~\cite{Ragavan2018}. They have been used to detect heme, a malaria biomarker, to a limit of detection~(LOD) of $2$~$\mu$M ($1.3$ $\mu$g/mL)~\cite{briand2012novel}. SPR optical biosensors have also been used to identify and characterize monoclonal and polyclonal antibodies of plasmodium falciparum histidine-rich~protein~2 (pfHRP-2), with a reported LOD of $0.517$~nM~\cite{sikarwar2014surface}. Recently, a graphene-enhanced copper SPR biosensor for detecting malaria DNA sequences has been reported, with an LOD in the pM range~\cite{wu2020ultrasensitive}. 

Planar SPR sensors, such as the ones described above, use propagating surface plasmon (SP) excitation to detect changes in the refractive index of the analyte due to molecular binding near the metal surface. Alternatively, structuring the metal surface into nanoparticles or patterned nanovoids enhances the electric field bound to the metal surface~\cite{couture2012eot}. Through bound SPs, nanostructured SPR sensors exhibit a narrow Fano resonance and a larger spectral shift per refractive index unit (RIU) compared to planar SPR sensors, due to the intensified sub-wavelength SP field confinement~\cite{couture2012eot, kim2011correlation, byun2007experimental}. These sensors are simple to fabricate and operate, and their optical properties, such as resonance frequency, can be tuned by adjusting parameters such as periodicity, hole size and shape, metal film thickness, and array lattice structure~\cite{henzie2009nanofabrication, haes2004nanoscale, prasad2019nanohole}. Bound SPs excited on periodic metallic nanostructures ($\textit{e.g.}$ plasmonic nanohole arrays and metasurfaces~\cite{li2023application}) are crucial to achieving ultrasensitive biomolecular detection~\cite{prasad2019nanohole, malic2007enhanced, anker2008biosensing}. The first experimental detection of the malaria biomarker plasmodium lactate dehydrogenase (pLDH) in whole-blood lysate using a nanostructured metasurface sensor was reported by Cho et al.~\cite{Cho2013}, using a gold film with a periodic nanohole array (NHA). More recently, Lenyk et al.~\cite{Bohdan2020} proposed a dual transducer malaria aptasensor using a gold NHA, combining electrochemical impedance spectroscopy (EIS) and SPR biosensing techniques in a single unit. An LOD of 23.5 nM was obtained based on the SPR sensing method. 

In addition to SPR techniques, other photonic studies have reported exploring possible sensors for the detection of malaria~\cite{Ragavan2018}. The best achieved LOD so far is $18$~fM, based on a plasmon-enhanced fluorescence immunosensor~\cite{Minopoli2020}, however, this approach requires highly specialized low-light level imaging equipment. Additional novel plasmonic-based configurations can be found in Refs.~\cite{kumar2024design, karki2024tuning, Shukla20,Shukla22}, while an overview of the various types of biosensors used for the detection of different malaria biomarkers is reported in Ref.~\cite{Krampa2020}. 

\begin{figure}[t]
\centering
{\includegraphics[width=0.45\textwidth]{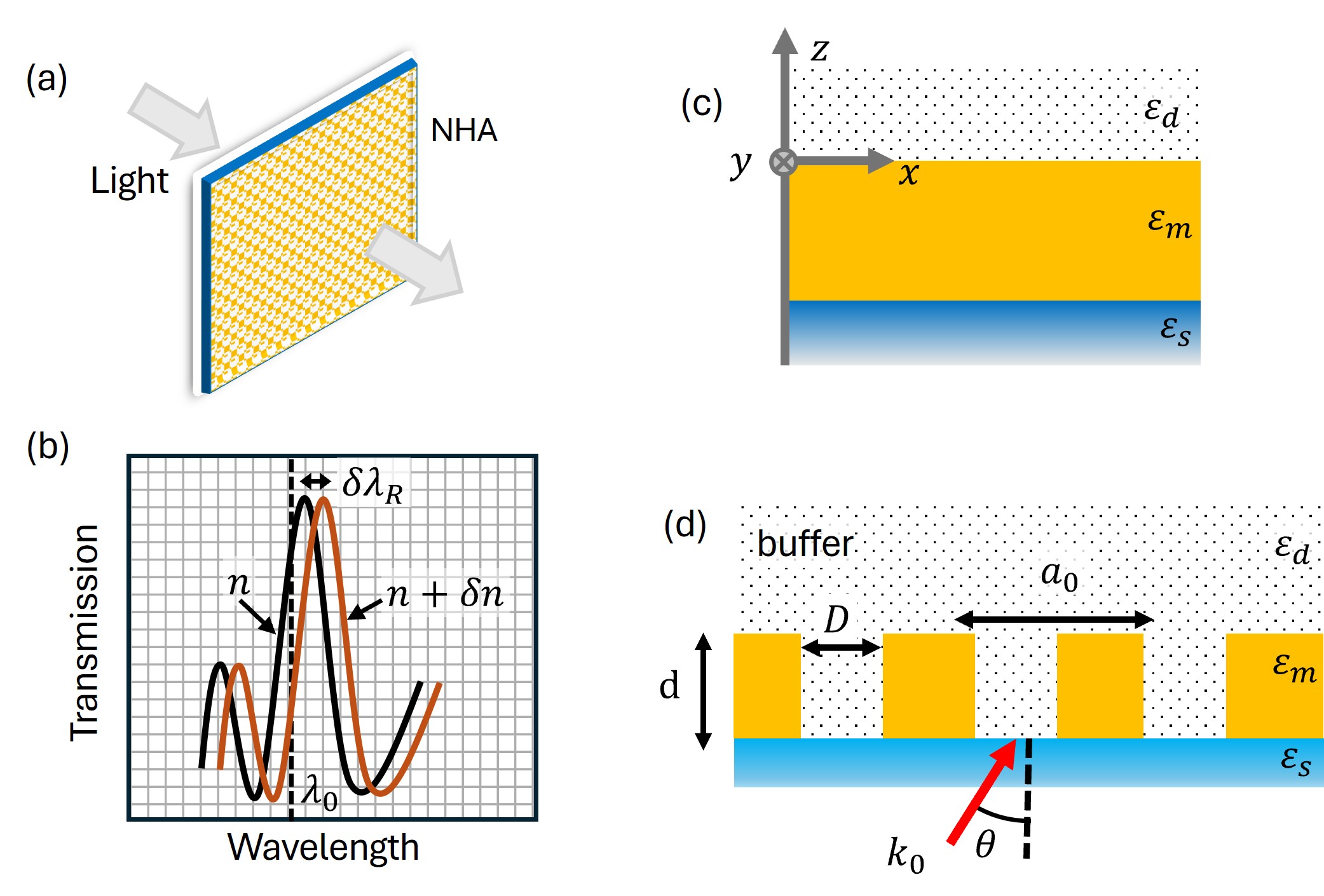}}
\caption{\label{spp_geometry} Formalism and working principle of a metasurface-based plasmonic biosensor. (a) A schematic drawing of the biosensor. Light is incident on the nanohole array (NHA) from the substrate side. (b) The spectra of the sensor before and after analyte binding. Plasmonic biosensing is achieved by monitoring the change in intensity or wavelength. Spectral sensing monitors the change of the transmission peak, $\delta \lambda_{R}$, when the refractive index $n$ is altered to $n+\delta n$, where $\delta n$ is the change or increase in the refractive index of the sensing medium. Intensity-based sensing monitors the change in transmittance, $\delta T$, at a fixed wavelength $\lambda_0$ off-resonance when $n$ changes to $n+\delta n$. (c) A three-layer SPR system that represents a typical configuration for exciting SPs. The parameters $\varepsilon_d,\;\varepsilon_m$, and $\varepsilon_s$, are the dielectric permittivity of the superstrate, metal, and substrate medium, respectively. (d) A cross-section view of the metasurface in panel (a). $D$ is the hole diameter, $a_0$ is the period, $d$ is the hole depth, $k_0$ is the free-space wavenumber of the incident wave and $\theta$ is the incidence angle.}
\end{figure}

In this work, we study theoretically the detection of the malaria biomarker pLDH by developing an effective microscopic and generic model. This allows us to predict the optical response of a metasurface-based plasmonic sensor for malaria detection and assess its potential LOD. The working operation of the sensor is summarized in Fig.~\ref{spp_geometry}(a) and (b). The model is based on a microscopic formalism previously reported in Ref.~\cite{Jung1998}, which we have adapted and developed further. By linking the refractive index of the bound pLDH protein to the bulk protein concentration in the buffer, our updated model allows one to estimate the effective refractive index of the bound pLDH near the sensor surface based on the concentration of pLDH in the buffer. 
We then show how this information can be combined with a knowledge of the bulk sensitivity of the non-functionalized sensor (which is easily measured in an experiment or numerically simulated) to predict the response of the sensor once it is functionalized. This enables a prediction of the LOD for the functionalized sensor based on a knowledge of only a handful of key quantities, importantly without the need for complicated functionalized experiments or simulations. 

In contrast to Ref.~\cite{Jung1998}, which models the sensor response based on SP field decay without considering binding kinetics, our approach develops the model further by combining SP field decay, Maxwell Garnett effective medium theory and Langmuir kinetics. This provides a unified approach that enables one to predict the sensor response from the bulk concentration, thereby facilitating sensor design and optimization. Furthermore, differently to Ref.~\cite{Cho2013}, which focused on the experimental detection of pLDH and a basic bulk sensitivity analysis, here our model is used not only to predict the LOD of a metasurface sensor for malaria as an example, but it has the potential to help guide the design of more optimal and reliable metasurface-based SPR biosensors in a broader context. The predictive advantage of our model is that it can provide the achievable LOD for a particular sensor using a small set of known parameters, giving a useful tool for sensor optimisation. 

In our study, we use optical simulations based on the finite element method (FEM) to calculate transmission spectra that are combined with a realistic light source spectrum to give a prediction for the expected experimental behavior of the model. While the developed model focuses on the surface sensing of pLDH spiked in phosphate-buffered saline (PBS), this approximates the experimental sensing of the protein in the more clinical setting of whole-blood lysate~\cite{lee2012highly, lee2014cationic, vazquez2018enzyme}. Based on the results, our effective model predicts that an LOD in the range of pM to nM for pLDH detection is achievable using a nanostructured gold metasurface. This level of sensing performance is competitive with current state-of-the-art plasmonic biosensors, but with the added benefit that specialized equipment is not required, due to the use of an uncomplicated transmission-based spectroscopy setup for probing the metasurface. This makes the example sensor we study using the effective model more practical and portable for point-of-care diagnostics. 

We also emphasize that it is not the low LOD achieved that is of significance in our work, but the theoretical model we develop that has practical utility, in the sense that it can potentially help future metasurface-based sensors improve their LOD values through design, prediction and testing stages. Our model may be used in the development of sensors of other types of malaria biomarkers, as well as biomarkers of other diseases, making the model's predictions useful in more broader applications. 

The paper is structured as follows. In Section II, we provide a~brief theoretical background of the sensor and describe the physics of the metasurface, accounting for the SP contribution to the calculated transmission spectrum. In Section III, we introduce the effective model, and calculate the sensitivity and LOD~based on spectral and intensity interrogation. In Section IV, we discuss our results, in Section V we highlight practical considerations of the model, and in Section VI we provide a summary and outlook.

\section{Theoretical background}
An evanescent surface wave is excited at a metal-dielectric interface when a TM-polarized optical wave is incident on the metal surface and momentum-matching conditions are met. The generated wave is called an SP excitation. It is due to the coupling of light to plasmons supported by the free conduction electron density of the metal surface~\cite{Maier2007}. To build up our effective model for the malaria biosensor we start by considering a simplified SPR geometry of a three-layer system consisting of a thin metal layer $\varepsilon_m$ ($\textit{e.g.}$ gold) of known thickness sandwiched between two thick layers--a substrate film $\varepsilon_s$ ($\textit{e.g.}$ glass), and a dielectric film $\varepsilon_d$ ($\textit{e.g.}$ PBS or analyte solution), as shown in Fig.~\ref{spp_geometry}(c). The dispersion relation of SPs at the top metal-dielectric interface is given as (see  Ref.~\cite{Maier2007})
\begin{equation}
k_{sp} = k_0 \sqrt{\frac{\varepsilon_d\varepsilon_m}{\varepsilon_d +\varepsilon_m}},
\label{resonant1}
\end{equation}
where $k_{sp}$ is the propagation constant of the SP, $k_0 = 2\pi/\lambda$ is the wavenumber of light in free space and $\lambda$ is the corresponding wavelength. SPs are laterally confined along the metal-dielectric interface and their field exponentially decays away from the interface. The decay length, also known as penetration depth, $\delta_m$ ($\delta_d$) of SPs in the metal (dielectric) is given by $1/k^m_z$ ($1/k^d_z$)~\cite{Maier2007,zhang2012surface}, where $k^m_z$ and $k^d_z$ are the components of the SP wavevector perpendicular
to the interface. SPs are very sensitive to small changes in the dielectric permittivity, $\varepsilon_d$, close to the metal surface~\cite{Homola2008, Homola2006}. However, above $\delta_d \sim 0.37 \lambda$, the sensitivity of the sensor is reduced~significantly~\cite{Jung1998}.\\

Metasurfaces are different from the flat surface considered above and comprise a two-dimensional array of sub-wavelength scatterers. Perforating a metal surface with periodic nanoholes creates the sub-wavelength scatterers. Our model considers a transmission metasurface, which is a variation of a multilayer system consisting of a perforated metal film between two thick dielectric films, with dielectric constants $\varepsilon_d$ and $\varepsilon_s$, as shown in Fig.~\ref{spp_geometry}(d). In such a system, each interface, such as the superstrate-metal interface ($\textit{e.g.}$ PBS-gold) or the substrate-metal interface ($\textit{e.g.}$ glass-gold) supports bound SPs (see Ref.~\cite{Maier2007}). The in-plane component of the wavevector of the incident light, $k_{||} = k_0n_s\sin\theta$ (with $n_s = \sqrt{\varepsilon_s}$ and $\theta$ the incidence angle), is enhanced by the grating factor $G = 2\pi/a_0$, where $a_0$ is the NHA period. The excitation of bound SPs is then achieved according to the phase matching condition $\vec{k}_{sp} = \vec{k}_{||}+nG\vec{x}+mG\vec{y}$, where $n$ and $m$ are integers associated with diffraction orders in the $x$ and $y$ directions~\cite{Maier2007}. The excited bound SPs at the two interfaces couple into the nanoholes and free space. They interfere with incident light that directly couples into the nanoholes to create extraordinary optical transmission (EOT)~\cite{Popov2000, Moreno2001}, where the transmission is larger than that predicted by standard aperture theory. At~normal incidence, $\theta = 0$, the spectral position of a resonant wavelength, $\lambda_{R}$, corresponding to a peak in the transmitted light through the metasurface is given by~\cite{Ghaemi1998, Ebbesen1998}
\begin{equation}
\lambda_{R} = \frac{a_0}{\sqrt{n^2+m^2}}\sqrt{\frac{\varepsilon_i\varepsilon_m}{\varepsilon_i +\varepsilon_m}},
\label{resonant2}
\end{equation}
where $i=d$ or $s$ (for the dielectric or substrate medium), allows for resonances associated with the dielectric-metal or substrate-metal interface, respectively, as shown in Fig.~\ref{spp_geometry}(d).
\begin{figure}[t]
\centering
\includegraphics[width=0.45\textwidth]{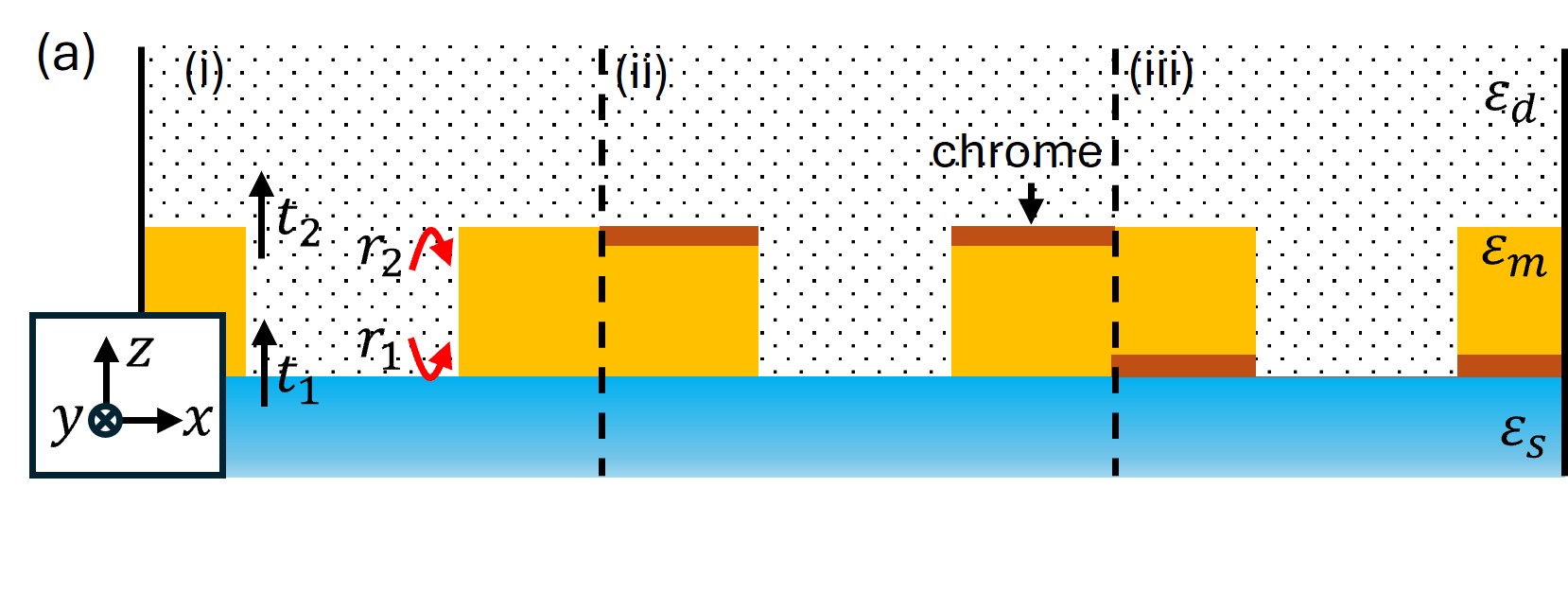}
\includegraphics[width=9cm]{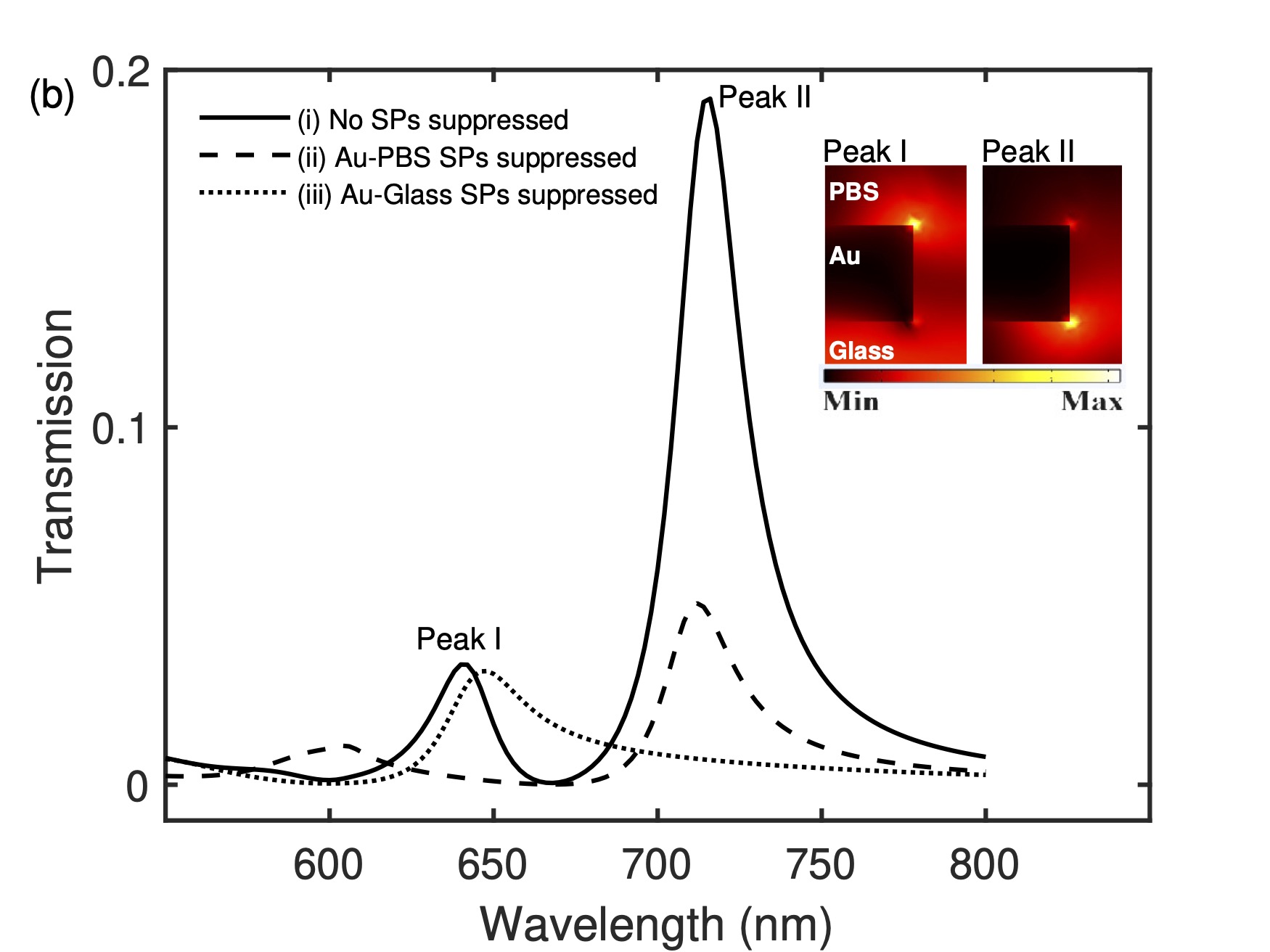}
\caption{Contribution of SPs to the transmission spectrum of the metasurface with hole diameter $D = 150$~nm, period $a_0 = 400$~nm and gold thickness $t = 150$~nm in phosphate-buffered saline (PBS). (a) The simulated geometries used. (i) The bare metasurface in PBS with no chrome added. (ii) A chrome layer ($20$~nm thick) is added between the gold-PBS interface. (iii) A chrome layer ($20$~nm thick) is added between the gold-glass interface. The parameters $t_1$ and $r_1$ (resp. $t_2$ and $r_2$) are the transmission and reflection coefficients of the fundamental Bloch mode of the array at the $\varepsilon_s$ (resp. $\varepsilon_d$) interface. (b) Simulated transmission spectra for the metasurface using COMSOL multiphysics software. Transmission spectra of the three cases. (i) A bare metasurface in PBS solution, where no SPs are suppressed. (ii) The SPs at the top dielectric interface are suppressed and, (iii) The SPs at the bottom glass interface are suppressed. The inset shows the electric field distribution within the hole (a cross-section view of a quarter cell). The maximum electric field enhancement $|E|/|E_0|$, at resonant wavelengths $\lambda_R = 650$~nm for peak I and $\lambda_R = 715$~nm for peak II, is 6 and 21, respectively. The minimum $|E|/|E_0|$ is zero. $|E|$ is the electric field norm. The dielectric constant of the substrate $\varepsilon_s$ and superstrate $\varepsilon_d$ are given in terms of the refractive indices $n_s = 1.52$ and $n_d = 1.335$, respectively. The permittivity of gold $\varepsilon_m(\lambda)$ included in the simulation was based on a Lorentz-Drude model~\cite{rakic1998optical}, taken from the \mbox{built-in} COMSOL optical materials database. A TM plane wave polarized in the $x$ direction propagating along $z$ was incident on the metasurface from the substrate side.} 
\label{sp_spectra_study}
\end{figure}

The EOT phenomenon was first reported by Ebbesen et al.~\cite{Ebbesen1998} in the near infra-red,  where it was attributed to the 
Wood anomalies of the metal. After extensive studies, the EOT effect is now interpreted as the result of hybridization between surface modes such as quasi-cylindrical waves (QCWs)~\cite{Beijnum2012} and SPs, where the least‐attenuated fundamental mode (\textit{i.e.} the SP mode) mediates the transmission of light through a Fabry-Perot type mechanism~\cite{Liu2010,lalanne2009, Liu2008}. On one side of the metal array excited SPs funnel light into the holes, and the hole modes bounce back and forth to create a resonant mode which interferes with incoming light that is directly coupled into the holes. More explicitly, for an incident plane wave, the free-space mode is coupled into an SP or couples directly into a hole. The SP mode is either transmitted or reflected within the plane of the array ($x-y$ plane), or coupled into the hole, or scattered back to free space. Each nanohole acts as a Fabry-Perot resonator, where light launched into the resonator, via an SP or directly, interferes with the light circulating in the resonator. At a specific wavelength $\lambda_R$, constructive interference occurs and results in an enhancement of the transmitted light~\cite{Beijnum2012}. Further details on the EOT phenomenon can be found in Refs.~\cite{Beijnum2012, Popov2000, Moreno2001,lezec2004diffracted}. 

The mechanism outlined above describes the main physics governing the metasurface transmission spectra and its resonance peaks that can be used for sensing. We now briefly discuss the contribution to EOT of the dominant SP modes by suppressing the SP mode on each interface, as shown in Fig.~\ref{sp_spectra_study}(a). To do this, we consider gold for the metal and adapt the model developed in Ref.~\cite{Beijnum2012}.

When the metasurface (bare-gold nanohole array, Fig.~\ref{sp_spectra_study}(a)(i)) is probed from the substrate (glass) side with TM-polarized light at normal incidence, it presents two EOT peaks. Peak I is around $645$~nm and peak II is around $715$~nm. The zeroth-order transmission, $T_m$, through the nanohole metasurface, numerically solved using the wave optics module of the FEM software COMSOL, is presented in Fig.~\ref{sp_spectra_study}(b) as a solid black line~labeled (i). Due to the rotational symmetry of the nanohole, a quarter of the unit cell was simulated to reduce computational time. The nanohole periodic lattice structure was maintained through perfect electric conductor and perfect magnetic conductor boundary conditions. Data from Rakic et al.~\cite{rakic1998optical} was used to
include the frequency-dependent permittivity, $\varepsilon_m$, of gold. The refractive indices for the glass substrate and PBS were
taken as 1.52 and 1.335, respectively. To avoid undesired numerical reflections back into
the interior of the computational region, the top and bottom domains of the model were
terminated by strong absorbing perfectly matched layers (PML). 

The transmission, $T_m$, from the COMSOL simulation can be expressed by a Fabry-Perot type equation as~\cite{Beijnum2012, Liu2008}
\begin{equation}
T_m = \frac{n_s}{n_d}\left\vert \frac{t_{1}t_{2} e^{ik_{sp}\ d}}{1 - r_{1}r_{2}e^{2ik_{sp}\ d}}\right\vert^2 ,
\label{Fabry}
\end{equation}
where $n_s$ and  $n_d$ are the refractive indices of glass (substrate) and PBS (superstrate), respectively, and $d$ is the nanohole depth. The parameters $t_1$ and $r_1$ (resp. $t_2$ and $r_2$) are the transmission and reflection coefficients of the fundamental Bloch mode of the array at the $\varepsilon_s$ (resp. $\varepsilon_d$) interface. Here, for simplicity, we consider a pure SP model for the $t_i$ and $r_i$ coefficients, with $i=(1,2)$ such that SPs are the only surface mode carrying energy between the holes (more generally the $t_i$ and $r_i$ coefficients can also account for other type of surface modes such as QCWs). The $t_i$ and $r_i$ are functions of microscopic scattering parameters~\cite{Beijnum2012, Liu2010}, and when inserted into Eq.~\eqref{Fabry} can be used to analyze the peaks in Fig.~\ref{sp_spectra_study}(b).

The transmission efficiency around peak I is completely diminished (near zero) when the SP at the superstrate (PBS) interface is suppressed by a chrome layer (20 nm thick) placed between the PBS and gold surface at the top of the nanohole, as shown in Fig.~\ref{sp_spectra_study}(a)(ii). The corresponding spectrum is shown as the dashed line (ii) in Fig.~\ref{sp_spectra_study}(b). When the SP at the PBS interface is suppressed, $r_2$ in Eq. $\eqref{Fabry}$ is approximately zero, and thus $T_m$ is only determined by its numerator. Since there are no SPs at the PBS interface to carry energy between holes, $t_2$ does not vary much with $\lambda$ and is reduced in magnitude compared to the case with no chrome~\cite{Beijnum2012, Liu2008}. However, due to SPs at the metal-glass interface, $t_1$ is dependent on $\lambda$. The peak that remains around $\lambda = 700 - 725$~nm, ($i.e.$ peak II) is therefore due to SPs at the metal-glass interface. 

On the other hand, adding chrome between the gold and substrate (glass) suppresses SPs at the glass-metal interface, as shown in~Fig.~\ref{sp_spectra_study}(a)(iii). The resulting spectrum, $T_m$, shown as a dotted line in Fig.~\ref{sp_spectra_study}(b), is thus mainly due to SPs at the metal-PBS interface. The peak around $\lambda = 600-650 $~nm (peak I) is therefore due to SPs at the metal-PBS interface. However, the chrome layer does not suppress to zero the $r_1$ coefficient due to the different refractive indices of the substrate and superstrate. Therefore, the transmission, $T_m$, is determined by both its numerator and denominator terms. The suppression of the numerator term due to $t_1$ ($\lambda$ independent) from the chrome layer is complemented by the enhancement of the denominator term through $r_1$ ($\lambda$ independent). This explains the nearly equal magnitude of the ``no SPs suppressed'' and the ``Au-Glass SPs suppressed'' transmission efficiency around peak I shown in Fig.~\ref{sp_spectra_study}(b). The inset in Fig.~\ref{sp_spectra_study}(b) confirms this picture and shows the near-field distribution around the resonance wavelengths (peaks) when there is no chrome present. For peak I, the field is more concentrated at the top corner end of the nanohole with a maximum field enhancement $|E|/|E_0| = 6$, where $|E_0|$ is the magnitude of the incident electric field norm. For peak II, the field is concentrated at the bottom corner with a maximum field enhancement $|E|/|E_0| = 21$. The minimum field norm for both peaks is zero. From the above analysis, it is clear that SPs are the main contributor to the EOT effect of peaks I and II~\cite{degiron2005role, Beijnum2012, Liu2008}. Peak I is used as the sensing peak in our model as it is expected to be more sensitive to changes in the dielectric properties at the metal-PBS interface.
\begin{figure*}[t]
{\includegraphics[width=0.7\textwidth]{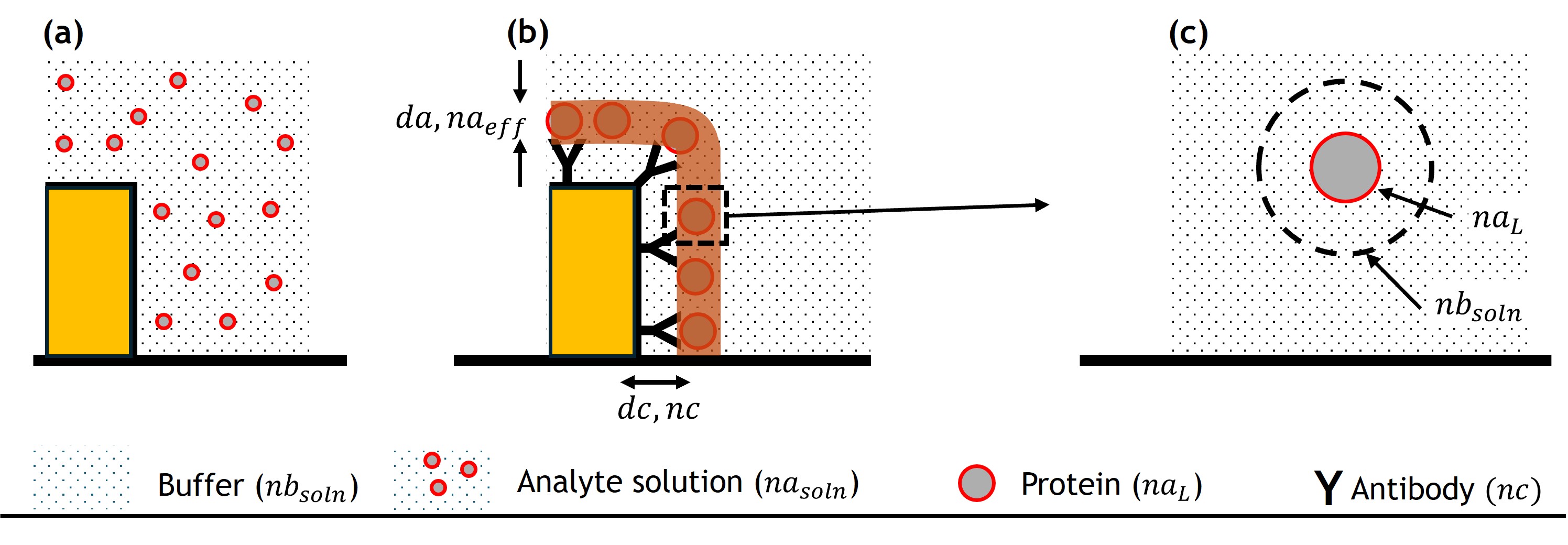}}
\caption{\label{sensor_model} Microscopic model formalism using Maxwell Garnett effective medium theory. (a) The non-functionalized metasurface with analyte solution. Proteins are uniformly distributed throughout the buffer medium. (b) Antibody-functionalized metasurface. Proteins bind to the immobilized antibodies forming a layer with thickness $da$ and effective refractive index $na_{\text{eff}}$. The refractive index $na_{\text{eff}}$ depends on the number of bound proteins. (c) A zoomed-in sketch of a bound protein shows the corresponding unit
cell used to derive the effective refractive index $na_{\text{eff}}$ using Maxwell Garnett theory. A dispersed phase ($\textit{e.g.}$ protein, $na_{\text{L}}$) is embedded in a continuous phase ($\textit{e.g.}$ PBS buffer solution, $nb_{\text{soln}}$).}
\end{figure*}

Assuming that the induced electromagnetic field $E_z~=~E_0e^{-z/\delta_d}$ due to bound SPs decays normal~to~the metasurface, with $\delta_d \sim 100 - 150$~nm for visible wavelengths~\cite{degiron2005role, yang2023sensitivity}, the surface functionalization matrix thickness usually needs to be optimized so that the analyte of interest is within the probe depth ($\delta_d$) to achieve desired surface sensing properties, such as sensor sensitivity and LOD. We note that a similar transmission metasurface has been discussed in Ref.~\cite{wu2011reflection} and its application in sensing has been reported in Ref.~\cite{Ding2015}. In Ref.~\cite{Ding2015}, the metasurface was used as an SPR sensor to demonstrate a lab-based bioanalyte detection of a cardiac troponin, a biomarker for a heart attack (myocardial infarction). The sensor achieved an LOD of $0.55$~ng/ml in serum~\cite{Ding2015}. Here, differently to this work, we develop an effective microscopic model for the metasurface as a more general SPR biosensor. This allows for a better optimization of sensing properties for any other SPR sensor based on the metasurface. We highlight the benefits of using the effective model for the case of sensing the malaria biomarker pLDH spiked in PBS buffer solution, which is an approximation to pLDH in the more relevant clinical setting of whole blood lysate.\\

\section{The microscopic model formulation}
\subsection{Non-functionalized and functionalized sensor}
 We now focus on the quantitative formalism of our metasurface sensor in terms of its microscopic architecture and the sensing parameters--sensitivity and LOD. We first discuss the influence of sensor functionalization in the overall sensing process. In Fig.~\ref{sensor_model}(a) we show a non-functionalized sensor (half cell) filled with analyte solution. The refractive index of the bulk analyte solution $na_{\text{soln}}$ can be estimated from~\cite{hand1935refractivity}
\begin{equation}
na_{\text{soln}} = nb_{\text{soln}} + \alpha L_0,
\label{non_fsensor}
\end{equation}
where $nb_{\text{soln}}$ is the refractive index of the buffer solution, $\alpha~=~\frac{dna_{\text{soln}}}{dL_0}$~is~the specific refractive index increment of protein and $L_0$ is the concentration of the protein measured in terms of mass per unit volume, \textit{i.e.} gram per $100$ mL of the buffer solution. The typical values of $\alpha$ for most proteins range from $0.175-0.230$ mL/g~\cite{hand1935refractivity,ball1998buffer}. For proteins from living cells it is approximately $0.185$ mL/g~\cite{barer1954refractometry, pierscionek1987refractive}.
\begin{table*}[ht]
\caption{\label{tab:table1}Definition of refractive index parameters used in the model.}
\begin{ruledtabular}
\begin{tabular}{ccccc}
 Type&Parameter&Fixed&Varies&Quantity\\ 
 \hline
 \multirow{3}{6em}{Bulk sensing}
 &$n$ &  & yes & general refractive index parameter\\
&$n_{s}$ & 1.52 &  & refractive index of the substrate medium\\
&$n_{d}$ &  & yes & refractive index of the superstrate medium\\
\hline
 \multirow{8}{6em}{Local sensing}
&$nb_{\text{soln}}$ & 1.335 &  & refractive index of the PBS buffer solution\\
&$na_{\text{L}}$ & 1.45 &  & refractive index of pure protein in aqueous state\\
&$na_{\text{soln}}$ &  & $L_0$ dependent & refractive index of the buffer + proteins solution\\
&$nc$ & 1.45 &  & refractive index of the biochemistry matrix (DSP + protein A + IgG)\\
 &$na_{\text{eff}}$ &  & $L_0$ dependent  & effective refractive index of the protein layer\\
&$n_{\text{eff}}$ &  & yes & effective refractive index of IgG, protein + buffer layer\\
&$n^0_{\text{eff}}$ & yes &  & effective refractive index of the biochemistry layer + buffer (no protein)\\
&$n(z)$ &  & yes & depth dependent refractive index of IgG, bound protein and buffer layers\\
\end{tabular}
\end{ruledtabular}
\label{Table1}
\end{table*}

When the metasurface is not functionalized, analytes (proteins) are uniformly distributed in the buffer solution. The refractive index of the protein solution $na_{\text{soln}}$ with low concentration can not be distinguished easily from that of the buffer solution. For example, based on Eq.~\eqref{non_fsensor}, the refractive index of $250$~nM of pLDH (MW $= 35$ kDa, $\alpha = 0.185$ mL/g) prepared in PBS buffer solution ($nb_{\text{soln}} = 1.335$) will be different to the refractive index of the buffer by an order of $10^{-4}$ RIU. This difference, $na_{\text{soln}}-nb_{\text{soln}}$, is small and hard to detect. For example, for a metasurface with a modest spectral sensitivity of $360$~nm/RIU, a $250$~nM solution will result in $\sim 0.05$~nm shift in the resonance peak. Such a small spectral shift cannot be easily resolved by standard spectrometers. The metasurface therefore needs to be functionalized with receptors to be able to detect a low analyte concentration.\\

Most malaria biomarkers are intracellular proteins~\cite{Jain2014}, and detecting them requires lysing the infected erythrocytes, a process that ends up diluting the whole-blood sample collected. Therefore a low concentration of protein is always presented in the analyte sample in most biosensing cases. As a result, a well-functionalized sensor surface is necessary to reduce biofouling, increase sensor sensitivity, and enhance binding specificity. In Fig.~\ref{sensor_model}(b) we depict a functionalized metasurface for plasmonic sensing of pLDH. As pLDH binds to immobilized antibodies, a pLDH layer is formed within the SP field. The protein solution $na_{\text{soln}}$ forms a gradient with a high effective refractive index close to the metal surface due to bound pLDH and a low refractive index layer,~$\sim nb_{\text{soln}} = 1.335$, away from the sensor surface. This is the main reason why a functionalized sensor becomes more sensitive than a non-functionalized one. 

Following Linda et al.~\cite{Jung1998}, the effective refractive index of the protein solution $n_{\text{eff}}$ can be thought of as a modified refractive index that the SP near field experiences. Neglecting the thickness of the functionalization layer ($i.e.$ immobilized receptor layer $dc$), $n_{\text{eff}}$ can be written as
\be
n_{\text{eff}} = \frac{2}{\delta_d}\int_{0}^{\infty} n(z)e^{-2z/\delta_d} \,dz\,,
\label{frist_model1}
\ee
where $n(z) = na_{\text{eff}}$ for $0<z\le da$ and equals to $nb_{\text{soln}}$ for $z>da$, as shown in Fig.~\ref{sensor_model}(b). However, the sensing region also involves an immobilized antibody layer of thickness $dc$ and refractive index $nc$, directly functionalizing
the metal surface. Therefore, using Eq. \eqref{frist_model1} and carrying out the integrals, a more accurate $n_{\text{eff}}$ that includes the receptor layer is given as 
\begin{eqnarray}
&n_{\text{eff}}=n_c\left(1-e^{-2dc/\delta_d}\right)+na_{\text{eff}}\left(1-e^{-2da/\delta_d}\right)e^{-2dc/\delta_d} \nonumber \\
&+nb_{\text{soln}}\left(e^{-2(dc+da)/\delta_d}\right). 
\label{frist_model11}
\end{eqnarray}
This equation, Eq.~\eqref{frist_model11}, takes into account how the refractive index of each layer contributes to the effective refractive index, $n_{\text{eff}}$, which governs the SPR response. For ease of reference, a summary of all the refractive indices used in our model is given in Table~\ref{Table1}. Among the terms in Eq.~\eqref{frist_model11}, the effective refractive index of the analyte layer $na_{\text{eff}}$ is the only unknown. However, it has been shown that $na_{\text{eff}}$ can be estimated from the analyte surface concentration using the size of the protein and its density, or the protein layer concentration~\cite{voros2004density, de1978ellipsometry}, an approach that is unfortunately not straightforward to carry out. Another simple effective medium model for plasmonic sensing was proposed by van de Hulst~\cite{hulst1981light, morales2017viability, reyes2018analytical} and applied in Ref.~\cite{morales2020plasmonic} to compare theoretical SPR results against experimental results. The effective refractive index of carbamazepine molecules ($n= 1.563$), a target analyte, deposited on a BSA coated gold thin film was calculated based on the molecule surface coverage and size parameter. \\

In this work, different to the above approaches, we instead link the effective refractive index $na_{\text{eff}}$ of the bound protein to the known bulk protein concentration~$L_0$. Using an effective medium theory, we combine the binding kinetics of the protein and the bulk protein concentration to obtain a value for $na_{\text{eff}}$. Our main assumptions are: (1) $na_{\text{eff}}$ is influenced by the bulk protein concentration~$L_0$, and (2) The binding affinity $K_A$ of the protein, which is the propensity with which proteins bind to the immobilized antibodies on the sensor, dictates the magnitude of the refractive index change. In the next subsection we discuss in detail this model.

\subsection{The Maxwell Garnett effective medium theory}
Consider the analyte, $\textit{e.g.}$ the pLDH protein spiked in a buffer solution, as a composite medium made up of two distinct materials, as shown in  Fig.~\ref{sensor_model}(c). The Maxwell Garnett (MG) effective medium theory describes the effective refractive index of a medium containing a dispersed phase (inclusions such as analyte molecules--pLDH) embedded in a continuous phase (host such as a buffer solution--PBS)~\cite{choy2015effective, markel2016introduction}. For small inclusions (low pLDH concentration), the effective refractive index of the analyte layer $na_{\text{eff}}$ can be found from \mbox{the~following~relation~\cite{niklasson1981effective, losurdo2013ellipsometry, choy2015effective, markel2016introduction}},
\begin{equation}
\frac{na^2_{\text{eff}} - nb^2_{\text{soln}}}{na^2_{\text{eff}} + 2nb^2_{\text{soln}}} =V_r\frac{na^2_{\text{L}} - nb^2_{\text{soln}}}{na^2_{\text{L}} + 2nb^2_{\text{soln}}},
\label{Maxwell-Garnett}
\end{equation}
where $na_{\text{L}}$ is the refractive index of the pLDH layer when the sensor is saturated--all available binding sites are occupied by proteins, $nb_{\text{soln}}$ is the refractive index of the bulk solution and $V_r$ is the volume fraction of the bound proteins to the total adsorbate layer volume. When the sensor is saturated we have $na_{\text{eff}} = na_{\text{L}} = 1.45$, which is the refractive index of the hydrated protein. The refractive index of most biological proteins is $1.45$~\cite{voros2004density}. Consider a simple adsorbate layer system where the host medium is a buffer solution, PBS, whose refractive index is $nb_{\text{soln}} = 1.335$ and $20\%$ of the total layer volume, $\textit{i.e.} V_r = 0.2$, is occupied by protein molecules with a refractive index of $na_{\text{L}} = 1.45$. The effective refractive index of such a layer according to Eq.~\eqref{Maxwell-Garnett} will then be $na_{\text{eff}}\sim 1.358$. On the other hand, when $V_r$ is zero, the layer contains no analyte and the effective refractive index will be that of PBS, \textit{i.e.} $na_{\text{eff}} = 1.335$. Alternatively, when $V_r$ is 1, the whole layer is occupied by analytes forming a closely packed protein layer, and $na_{\text{eff}}$ will be equal to $na_{\text{L}} = 1.45$.

\subsection{Fractional volume $V_r$}
We now turn our attention to the parameter $V_r$--the fractional volume occupied by the analyte in the adsorbate layer. The fractional volume is directly linked to the number of bound analytes, the dissociation rate, and the surface area of the analyte. In many kinetic controlled interactions where analytes such as pLDH bind to immobilized antibodies IgG, the sensor response is usually modeled using a Langmuir adsorption model~\cite{schasfoort2017handbook, Ayawei2017}. We start by assuming that one pLDH molecule binds to only one binding site of the IgG antibody. The system should reach equilibrium at the final stage of interaction, where the net effect of the association and dissociation process is zero. The equation describing this dynamic equilibrium~is~given~by~\cite{schasfoort2017handbook}
\be
\Theta_A = \frac{K_AL_0}{1 + K_AL_0},
\label{eq:Lmodel}
\ee
where $\Theta_A$ is the fractional number of binding sites occupied by proteins, $L_0$ is the concentration of protein in the buffer, and $K_A$ is the binding affinity. The parameter $\Theta_A$ is equivalent to $V_r$ because the percentage volume occupied by proteins depends on the number of proteins forming the thin adsorbate layer. Assuming a protein occupies all of the space above each antibody, we have $\Theta_A = V_r = L_0^{\text{ad}}/{B_{\text{max}}}$, where $L_0^{\text{ad}}$ is the concentration of bound protein in the analyte layer and $B_{\text{max}}$ is the maximum concentration of functional receptors immobilized on the surface. It is convenient to write Eq.~\eqref{eq:Lmodel} in terms of the equilibrium dissociation constant $K_D = 1/K_A$. Using this and the above information, we find that
\be
V_r = \frac{L_0^{\text{ad}}}{{B_{\text{max}}}}= \frac{L_0}{K_D+L_0}.
\label{eq:Lmodel2}
\ee
When the sensor response $\delta R$, \textit{e.g.} $\delta \lambda_R$ or $\delta T$, is written in terms of Eq.~\eqref{eq:Lmodel2}, it resembles the well-known Langmuir adsorption model~\cite{lavin2018determination}. That is
\be
\delta R = \frac{\delta R_{\text{max}}L_0}{K_D+L_0} = V_r\delta R_{\text{max}} .
\label{eq:Lmodel3}
\ee
When $V_r = 0$, we have $\delta R = 0$ and when $V_r=1$ then $\delta R = \delta R_\text{max}$. We combine the Langmuir model and MG theory to obtain the effective refractive index $na_{\text{eff}}$ of the thin-bound analyte layer at equilibrium. Using Eq.~\eqref{eq:Lmodel2}, our model equation, Eq.~\eqref{Maxwell-Garnett}, is then given by
\begin{equation}
\frac{na^2_{\text{eff}} - nb^2_{\text{soln}}}{na^2_{\text{eff}} + 2nb^2_{\text{soln}}} =\left(\frac{L_0}{K_D+L_0}\right)\frac{na^2_{\text{L}} - nb^2_{\text{soln}}}{na^2_{\text{L}} + 2nb^2_{\text{soln}}}.
\label{Maxwell-Garnett_new}
\end{equation}
While $V_r = 1$ represents a scenario where the entire layer volume is occupied by proteins, this is typically not achievable in practice due to the physical properties of proteins, the presence of the buffer, and the limitations of molecular packing~\cite{voros2004density}. We introduce a scaling parameter $f_{\text{max}}$, so that the corrected volume fraction $V^c_r$ is defined as  
\be
V^c_r = \Theta_A f_{\text{max}}= V_{r}f_{\text{max}} = \left(\frac{L_0}{K_D+L_0}\right)f_{\text{max}}.
\label{Maxwell-Garnett_final0}
\ee
In most cases, $f_{\text{max}}$ will be less than 1, reflecting a more realistic packing density and the presence of other components in the sensing layer. For example, in a well-packed monolayer of spherical proteins, $f_{\text{max}}$ is around 0.8-0.9 due to packing limitations ($\textit{e.g.}$ hexagonal close packing)~\cite{lu1998denaturation}. Large or irregularly shaped proteins will pack less efficiently, leading to lower values of $f_{\text{max}}$. In practice, $f_{\text{max}}$ would be the highest value that still yields a physically meaningful solution for $na_{\text{eff}}$ before the model becomes unrealistic $\textit{e.g.}$, when the calculated $na_{\text{eff}}$~no longer aligns with experimental observations. By modifying Eq.~\eqref{Maxwell-Garnett} with $V^c_r$ we find,
\begin{equation}
na^2_{\text{eff}} = nb^2_{\text{soln}}(1+2V^c_r\gamma)(1-V^c_r\gamma)^{-1},
\label{Maxwell-Garnett_final}
\end{equation}
where $\gamma = (na^2_L - nb^2_{\text{soln}})(na^2_L +2nb^2_{\text{soln}})^{-1}$. Eq.~\eqref{Maxwell-Garnett_final} can be simplified further into the form \textbf{$y = x+qx$} such that,
\begin{equation}
na^2_{\text{eff}} = nb_{\text{soln}}^2+\left[\frac{3V^c_r\gamma}{1-V^c_r\gamma}\right]nb_{\text{soln}}^2.
\label{Maxwell-Garnett_final_new}
\end{equation} 

We can now calculate the response of the SPR sensor (spectral or transmission) using $\delta R = S_R\delta n_{\text{eff}}$, where $S_R=\left|\frac{\delta R}{\delta n_{\text{eff}}}\right|$ is the sensitivity of the sensor response $R$ in terms of changes in the effective refractive index $n_{\text{eff}}$~\cite{Beijnum2012}. We have that $\delta n_{\text{eff}} = n_{\text{eff}}-n^0_{\text{eff}}$, where $n^0_{\text{eff}}$ is the value of $n_{\text{eff}}$ in Eq.~\eqref{frist_model11}, but with no analyte layer, \textit{i.e.} when $na_{\text{eff}}
 = nb_{\text{soln}}$. Using Eq.~\eqref{frist_model11} for $n_{\text{eff}}$ and $n^0_{\text{eff}}$, we find
 \begin{equation}
\delta R = S_R\delta n_{\text{eff}}= S_R(na_{\text{eff}}-nb_{\text{soln}})\left(1- e^{-2da/\delta_d}\right)e^{-2dc/\delta_d}.
\label{resonance_shift_estimate}
\end{equation}

In the above, $S_R$ is the same as the bulk sensitivity, $S_R = \frac{\delta R}{\delta n_d}$. The reason for the equivalence is that the refractive index, $n_{\text{eff}}$, is simply the effective refractive index above a non-functionalized sensor, $i.e.$ $n_{\text{eff}}=n_d$, that takes into account the presence of the antibody (functionalization layer), protein and buffer. We can therefore use Eq.~\eqref{resonance_shift_estimate} to predict the sensor response for a functionalized sensor from a knowledge of the bulk sensitivity, $S_R$, of the non-functionalized sensor. Bulk sensitivity considers changes in the refractive index, $n_d$, throughout the entire sensing volume. It can be found simply from measurements of the non-functionalized sensor or basic FEM modelling.\\

It should be noted that the bulk sensitivity, $S_R$, of a non-functionalized sensor is often quoted in literature as quantifying the sensing performance of the sensor, however, it does not give enough information to quantify the functionalized version's actual response to a targeted analyte. If $S_R$ is obtained experimentally (or via a basic FEM simulation), then with the knowledge of the antibody biochemistry layer size ($dc$), the protein size ($da$), the buffer solution ($nb_{\text{soln}}$) and $na_{\text{eff}}$ (through the effective microscopic model given in Eq.~\eqref{Maxwell-Garnett_final_new}), the functionalized sensor's actual response $\delta R$ can be predicted using Eq.~\eqref{resonance_shift_estimate}. The model relies on combining several core concepts from the literature, such as Maxwell Garnett theory, Langmuir adsorption and SPR-based metasurface sensing. As far as we are aware, combining these concepts to give such a predictive model has not been proposed in the literature. 

For instance, the work in Ref.~\cite{Cho2013} studied a gold metasurface for biosensing with an experimental focus, but did not provide a model to predict the functionalized sensor response and thus the LOD, only a measurement of the bulk sensitivity, $S_R$, and a few related measurements of selected biomarker concentrations for the functionalized sensor. On the other hand, in Ref.~\cite{Jung1998}, a theoretical model was developed to predict an SPR sensor's response based on a knowledge of the effective refractive index of the biomarker layer. However, the effective refractive index was not explicitly linked to the bulk concentration of the biomarker in the buffer and therefore the response of the sensor based on biomarker concentration cannot be predicted from this model. 

On the other hand, our model extends that of Ref.~\cite{Jung1998} by integrating Langmuir binding and Maxwell Garnett effective medium theory. This combination of concepts provides a framework based on Eq.~\eqref{resonance_shift_estimate} with practical utility that enables the response of a functionalized sensor to concentration changes in the biomarker to be predicted based on bulk sensing measurements (or basic FEM simulations) of the non-functionalized sensor. This is a clear practical advantage in sensing experiments -- the measurement of bulk sensitivity, $S_R$, is usually the first step in the characterisation of a sensor and the model requires a knowledge of only this parameter and a handful of other key quantities, without the need for complicated functionalized experiments or simulations. It is therefore a more complete and practically useful model than the one introduced in Ref.~\cite{Jung1998}.

When combining Eqs.~\eqref{Maxwell-Garnett_final_new}~and~\eqref{resonance_shift_estimate}, we see that there are mainly three factors that influence the magnitude of $\delta R$ due to changes in the effective refractive index of the bound protein: (1) The protein size ($da$) and its molecular weight--larger molecules generally produce a greater change in refractive index upon binding. (2) Binding affinity--the association rate constant $K_A$ and dissociation rate constant $K_D$ at equilibrium. Faster association rates result in a larger change in refractive index, hence stronger signals. (3) Sensor surface properties ($dc$ and $nc$). The characteristics of the immobilized receptor, such as density and orientation, can impact the binding kinetics and the resulting change in the effective refractive index $na_{\text{eff}}$, which then affects $\delta R$ and ultimately the sensor~sensitivity~\cite{makaraviciute2013site,vashist2011effect}. A thinner functionalization layer (i.e. $dc\ll \delta_d$) generally leads to a higher sensitivity as the protein layer is close to the SP field, which enhances the signal change.
\subsection{Simulation of the sensor model}
\begin{figure}[t]
\centering
\includegraphics[width=0.4\textwidth]{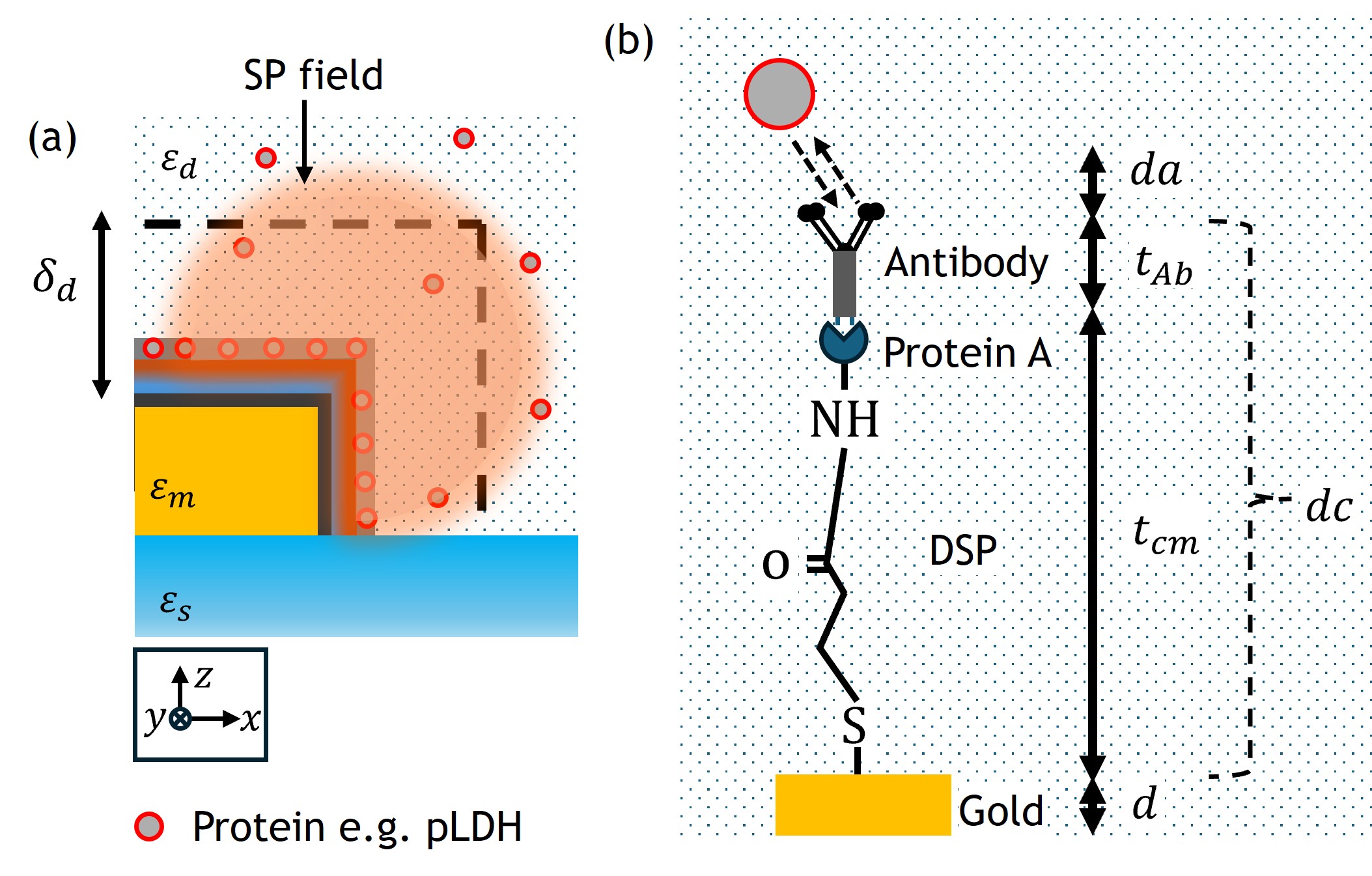}\
\includegraphics[width=0.4\textwidth]{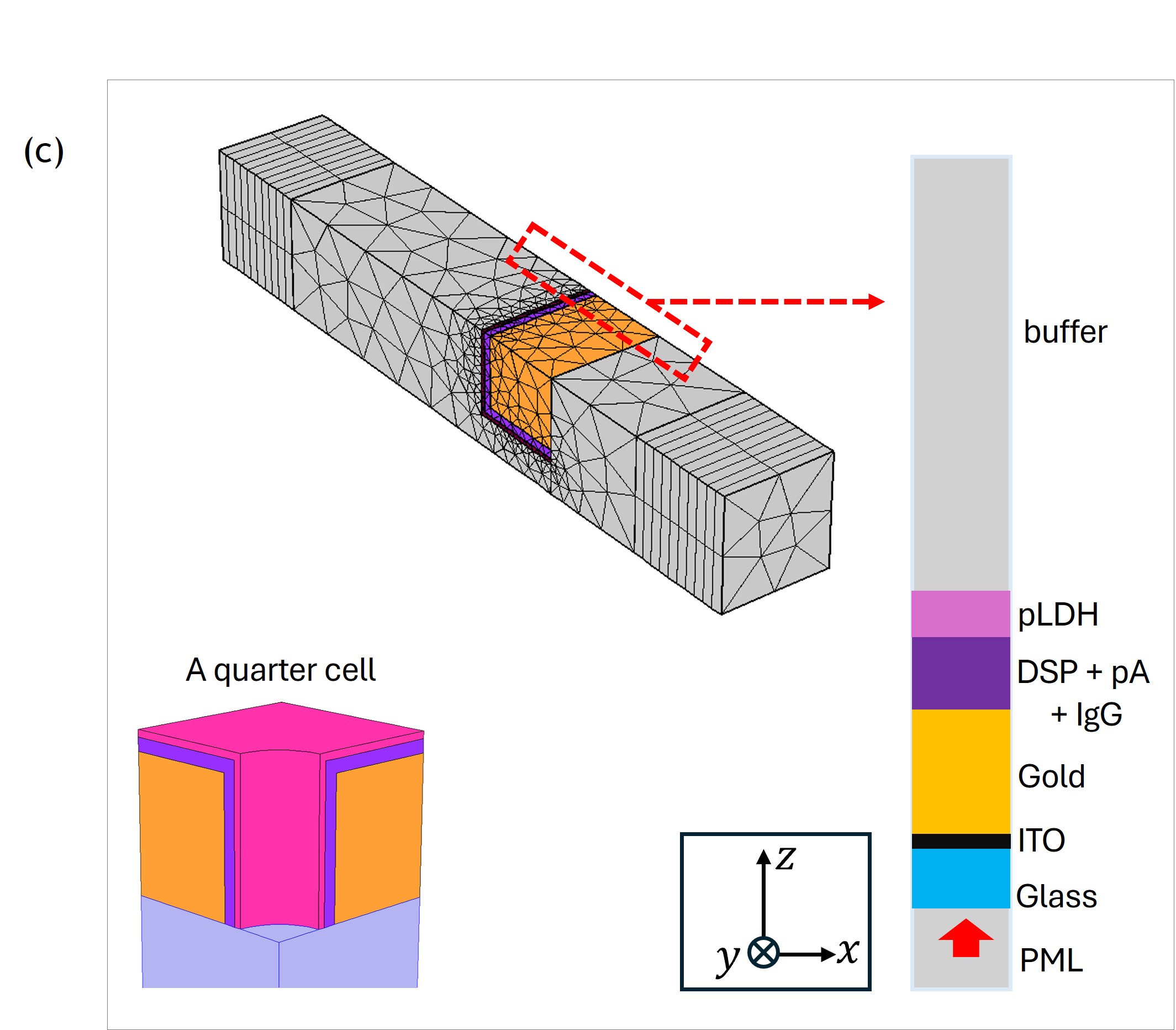}
\caption{\label{COMSOL_model} A schematic of the simulated model involving a quarter~of~a unit cell. (a) The cross-section view of our sensor, where $\delta_d$ is the~SP decay length. (b) A zoomed-in region of the sensor shows the surface chemistry and a binding event near the sensor surface. The biochemistry layer (DSP + Protein A) has a thickness $t_{cm} = 6$~nm, the antibody layer has a thickness $t_{Ab} = 8.4$~nm, and the analyte layer has a thickness $da = 7.9$~nm--the size of pLDH. Dashed arrows represent the kinetic interaction of the analyte to the immobilized antibody. Alternating dots represent the buffer solution with index $nb_{\text{soln}}$. (c) Adlayers are packed in series from the gold surface: DSP coating + protein A + IgG and a pLDH bound layer. Buffer solution covers the whole sensor surface. Gold is poorly
bound to glass, and so an adhesion layer of ITO ($5$~nm thick) is used. Glass with refractive index $n_s = 1.52$ is used as a substrate. Light (red arrow) enters the sensor from below.} 
\end{figure}
In order to check the correctness of our effective microscopic model for predicting the sensor response, based on Eq. \eqref{resonance_shift_estimate}, we simulated the functionalized metasurface sensor introduced in Section II using COMSOL. For both the microscopic model and the COMSOL simulation, we use the same $na_{\text{eff}}$ given by Eq. \eqref{Maxwell-Garnett_final_new}. To obtain the bulk sensitivity for the microscopic model, $S_R$, we simulate the non-functionalized metasurface in COMSOL, although this would ideally be obtained from an experiment. The microscopic architecture of the sensor used for simulating the sensing of pLDH spiked in PBS solution is presented in Fig.~\ref{COMSOL_model}(a) and (b), and the COMSOL geometry is shown in Fig.~\ref{COMSOL_model}(c). 

The surface functionalization strategy used in the design of the COMSOL model is reported in Ref.~\cite{Cho2013}. It involves site-directed immobilization of antibodies (IgG type) through protein A~\cite{schmid2006site}. Briefly, the gold surface is pre-coated with a dithiobis(succinimidyl propionate) (DSP) cross-linker, via a disulfide bond (S-S) formation between the thiol group (-SH) on the gold surface and the thiol group of the DSP molecule. Protein A is then incubated to bind to the DSP NHS-ester ends through covalent bonding~\cite{hermanson2013bioconjugate}. Protein A then captures the Fc region of the IgG-antibody exposing the binding sites of the antibody that allow for binding to pLDH. A layer of pLDH is formed on top of the antibodies during sensing. We model the surface chemistry as a single dielectric layer (\textit{i.e.} DSP + pA + IgG) with thickness $dc = 14.5 $~nm and refractive index $nc = 1.45$. The DSP forms a self-assembled monolayer on the gold surface that has an effective thickness of $1$~nm~\cite{grubor2004novel, Hermanson2013chap13} and refractive index of $n = 1.45$~\cite{schmid2006site}, which is the refractive index of most alkanethiols~\cite{peterlinz1996two, haes2004nanoscale}. Protein A is a single-domain protein with an overall size of about $5.0$~nm in diameter and $5.0$~nm in height~\cite{grubor2004novel}, with refractive index of $n=1.45$~\cite{tan2008nanoengineering}. The effective thickness of the antibody~(IgG) layer is taken as $8.4$~nm~\cite{tan2008nanoengineering}, and has a refractive index of~1.45.\\

The adlayers (DSP, protein A, antibody and pLDH layers) only enclose the gold surface of the nanohole and are not present at the bottom flat surface of the nanohole because, for DSP, sulfur does not easily bind to glass due to the absence of reactive amine groups on the glass substrate~\cite{Hermanson2013chap13}. Glass and silica are inert to sulfur. Their surfaces are primarily composed of silanol (Si–OH) groups, which do not react directly with the thiol group or N-hydroxysuccinimide (NHS) esters present in DSP~\cite{hermanson2013bioconjugate}. We also assume that during sensing, the bound pLDH thickness $da$ remains constant and is equal to the pLDH geometric thickness of $7.9$~nm (the size of the pLDH protein). Therefore, the only variable affecting $na_{\text{eff}}$ in the pLDH layer is the refractive index of bound proteins. The effective refractive index, $na_{\text{eff}}$, of the pLDH layer increases with an increase in the bulk pLDH concentration $L_0$ according to our effective microscopic model, as given in Eq.~\eqref{Maxwell-Garnett_final_new}. The saturation refractive index $na_{\text{L}}$ of the pLDH layer is chosen to be 1.45 RIU, \textit{i.e.} when all IgG binding sites are filled with pLDH and the IgG-pLDH interaction is at equilibrium. 

The precise $K_D$ value pertinent to the binding of IgG with pLDH remains inadequately documented in the literature. Here, a modest $K_D$ of 0.25 nM is chosen, which is within the range of a highly specific and optimized affinity. For instance, a value for $K_D$ of 0.306 nM has been reported for pfLDH binding to the IgG1 (10C4D5) antibody~\cite{kaushal2014production}. Nevertheless, the affinities of antibodies are contingent upon the specific epitope, plasmodium species (such as P. falciparum and P. vivax) and the conditions under which the assay is conducted~\cite{lee2011development, kaushal2014production, piper2011opportunities}. For natural anti-pLDH IgG binding to pLDH, a high $K_D$ of ten to hundreds of nM might even be expected~\cite{lee2011development, kaushal2014production}. 

For the sensor simulations, the optical response of the sensor is calculated using COMSOL. An electromagnetic (EM) wave linearly polarized along the $x$ direction is incident from the substrate side. The same model approach, as detailed in Section II, was followed.

\subsection{Sensitivity and Limit of Detection calculation}
The pLDH-IgG interaction is kinetically controlled and the Langmuir model that fully describes the sensor's optical response is given by Eq.~\eqref{eq:Lmodel3}, where $R$ is an explicit observable parameter (transmission $T$ or resonance wavelength $\lambda_R$) that changes with respect
to a change in the pLDH effective refractive index, $n_{\text{eff}}$. For $R=T$, the sensor sensitivity $S_T$ at a fixed wavelength $\lambda_0$ (off-resonance) is given by~\cite{Homola2006}
\begin{equation}
S_{T} = \left\vert\frac{\delta T}{\delta n_{\text{eff}}}\right\vert_{\lambda_0},
\label{sensitivity1}
\end{equation}
where $n_{\text{eff}}$ is defined according to Eq.~\eqref{frist_model11} and we assume a linear response. For $R=\lambda_R$, the formula can be used for wavelength interrogation by putting $\lambda_R$ in place of $T$. The spectral sensor sensitivity $S_{\lambda_R}$ is therefore given as~\cite{Homola2008, Homola2006}
\begin{equation}
S_{\lambda_R} = \frac{\delta\lambda_R}{\delta n_{\text{eff}}}.
\label{sensitivity}
\end{equation}
We define the LOD as the minimum detectable change of the effective refractive index, $\Delta n_{\text{eff}}$ (or protein concentration $\Delta L_0$), that the sensor can resolve. The LOD for a particular quantity ($n_{\text{eff}}$ or $L_0$) is the ratio of signal noise, $\Delta R$, to the sensor sensitivity $S_{R}$ for that quantity. For $n_{\text{eff}}$, it is written as~\cite{Homola2008, Homola2006, Lee2021}
\begin{equation}
\textrm{LOD} = \Delta n_{\text{eff}}=  \frac{\Delta R}{S_{R}}.
\label{LOD}
\end{equation}
Here, $\Delta R$ is the noise associated with the measurement of $R$, which is the standard deviation of the sensor response--usually extracted from several responses of a blank solution with $L_0=0$. 

In this theory work, as an example, noise is introduced from a knowledge of $\Delta T$ originating from a realistic white light source. For a given $\Delta T$, the spectral noise $\Delta\lambda_R$ can then be approximated as~\cite{mekid2008propagation}
\begin{equation}
\Delta\lambda_R \simeq {\left\vert\frac{dT}{d\lambda}\right\vert_{\lambda_0}}^{-1}\Delta T = {\left\vert\frac{dT}{dn_{\text{eff}}}\right\vert_{\lambda_0}}^{-1}\frac{d\lambda}{dn_{\text{eff}}}\Delta T,
\label{delta_lambda}
\end{equation}
where ${\left\vert\frac{dT}{d\lambda}\right\vert_{\lambda_0}}$ is the gradient of the spectral profile at a fixed off-resonance wavelength $\lambda_0$, ${\left\vert\frac{dT}{dn_{\text{eff}}}\right\vert_{\lambda_0}}$is given by $S_T$ in Eq.~\eqref{sensitivity1}, which describes the sensor's transmission sensitivity, and $\frac{d\lambda}{dn_{\text{eff}}}$ is given by the spectral sensitivity $S_{\lambda_R}$ from Eq.~\eqref{sensitivity}. Here, as $\lambda_R$ corresponds to a peak in the transmission spectrum where $\frac{dT}{d\lambda} = 0$, strictly speaking we require the second-order term in the Taylor expansion of the error propagation formula~\cite{mekid2008propagation}. We use a wavelength $\lambda_0$ slightly off-resonant from $\lambda_R$ where the first-order term (linear) is used to obtain an approximation~of~$\Delta\lambda_R$.
\begin{figure*}[t]
\centering
\includegraphics[width=0.95\textwidth]{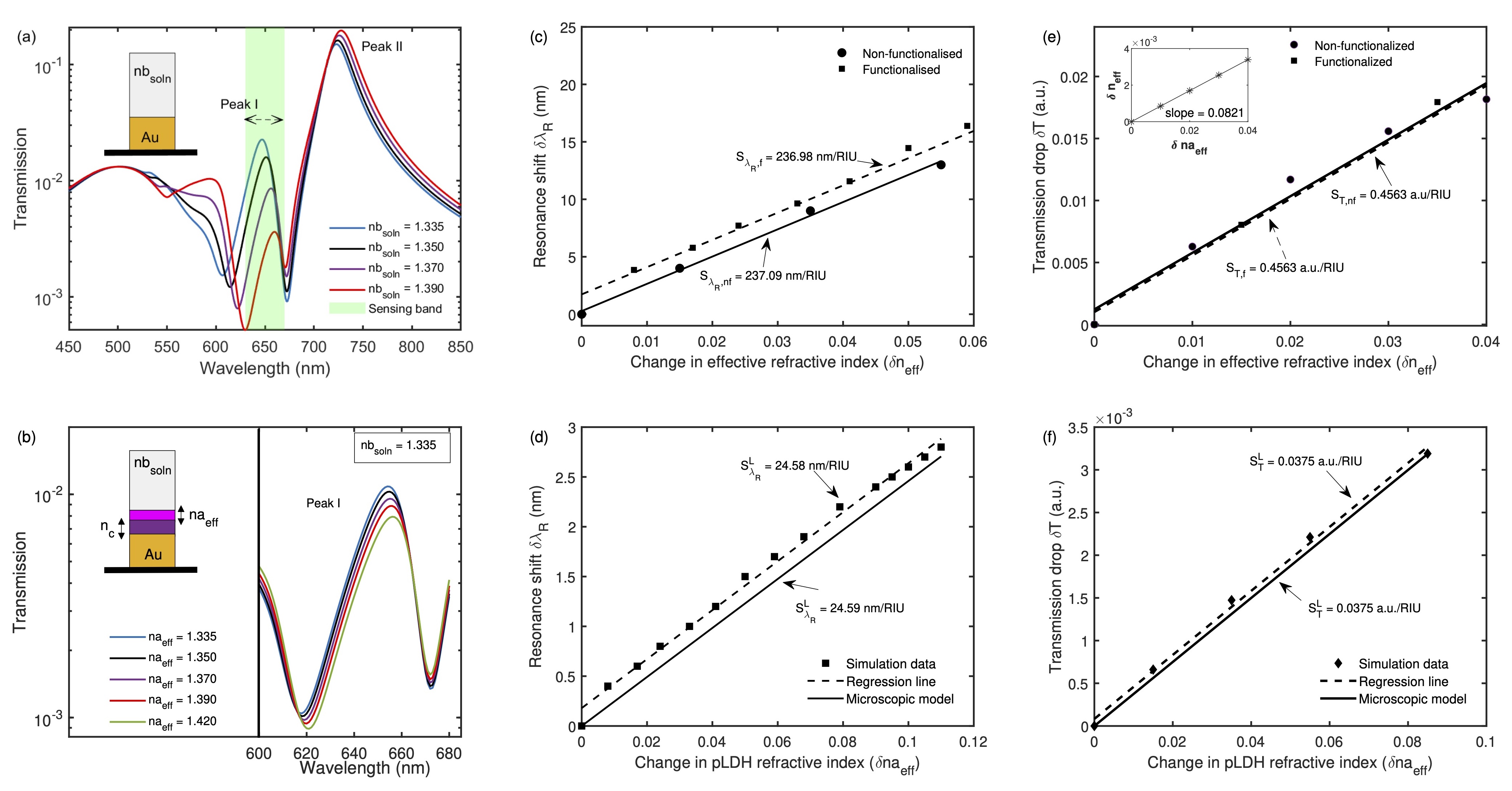}
\caption{\label{model_responce} Sensor spectral response and sensitivity. (a) Transmission spectra of the non-functionalized metasurface in different liquids, $nb_{\text{soln}} = 1.335$ (PBS buffer solution) to $1.390$. (b) Transmission spectra of the functionalized sensor where $na_{\text{eff}}$ is the refractive index of the pLDH layer bound to the antibody layer with $n_c = 1.45$. The inset shows the adsorbate layer architecture above the gold (Au) surface. (c) The bulk spectral-sensitivity of a functionalized sensor, $S_{\lambda_R,\;\text{f}}$, and non-functionalized sensor, $S_{\lambda_R,\;\text{nf}}$, are compared. (d) The calculated local sensitivity $S^L_{\lambda_R}$ using the simulated data in panel (b) and the microscopic model from Eq.~\eqref{resonance_shift_estimate}. (e) The bulk intensity-sensitivity of a functionalized sensor, $S_{T,\;\text{f}}$, and non-functionalized sensor, $S_{T,\;\text{nf}}$, are compared. (f) The calculated local sensitivity $S^L_{T}$ using the simulated data in panel (b) and the microscopic model from Eq.~\eqref{resonance_shift_estimate}. The $n_{\text{eff}}$ values in panels (c) and (e) are calculated from Eq.~\eqref{frist_model11}, where $na_{\text{eff}}$ was varied from~$1.335-1.375$. The inset in panel (e) shows how the change in the effective refractive index, $na_{\text{eff}}$, of the pLDH layer, affects the change in the effective refractive index,~$n_{\text{eff}}$, above the Au surface, as given by Eq.~\eqref{frist_model11}. The deviation in the $S_{\lambda_R}$ magnitude, in panel (c), of $237.09$~nm/RIU to $236.98$~nm/RIU for the non-functionalized and functionalized cases, respectively, is attributed to the small modification of probe depth, $\delta_d$, due to the presence of adsorbate layers in the functionalized sensor. To achieve optimal~fitting of the microscopic model in panels (d) and (f), the probe depth, $\delta_d$, was varied from $100-160$~nm, giving 108.9~nm and 151.7nm, repectively. Further details on sensor sensitivities are given in Appendix A, in Section 1.}
\end{figure*}

While Eq.~\eqref{LOD} can be used together with $\Delta T$ and $\Delta \lambda_R$ (from Eq.~\eqref{delta_lambda}) to obtain the theoretical LOD, it should be noted that experimentally the LOD is usually defined as LOD $=\sigma_R / S_R$, where $\sigma_R = 3\Delta R$, providing a $99\%$ confidence level. Here, $\Delta R$ is taken from blank measurements where $L_0= 0$ and $S_R = \frac{\delta R}{\delta L_0}$ is the slope of a calibration curve--a plot of $\delta R$ against $L_0$ with data fit to the Langmuir model. The intersection of the $\sigma_R$ line (also called the blank signal level) and the Langmuir fit determines the lowest concentration $L_0$ that can be resolved, $i.e.$ the LOD, which is $\Delta L_0 = \frac{\sigma_R}{S_R}$. The sensor response $\delta\lambda_R$ or $\delta T$ is not perfectly linear except for a small range of concentration $L_0$, \textit{i.e.} a weak protein concentration. Hence, $S_{R}$ is constant over only a narrow working range of $L_0$. Similar observations have also been reported in Refs.~\cite{Jung1998,liedberg1993principles}. The LOD analysis based on the experimental calibration curve method discussed above, $i.e.$ a 3:1 signal-to-noise ratio method, is more satisfying from a scientific standpoint, because the sensor response is linear at weak concentrations and within a few units of the LOD~\cite{lavin2018determination}. We therefore use the definition $\text{LOD}= {3\Delta R}/{S_R}$ in our analysis to connect the theoretical results to experimental convention. 
\begin{figure*}[t]
\centering
\includegraphics[width=0.75\textwidth]{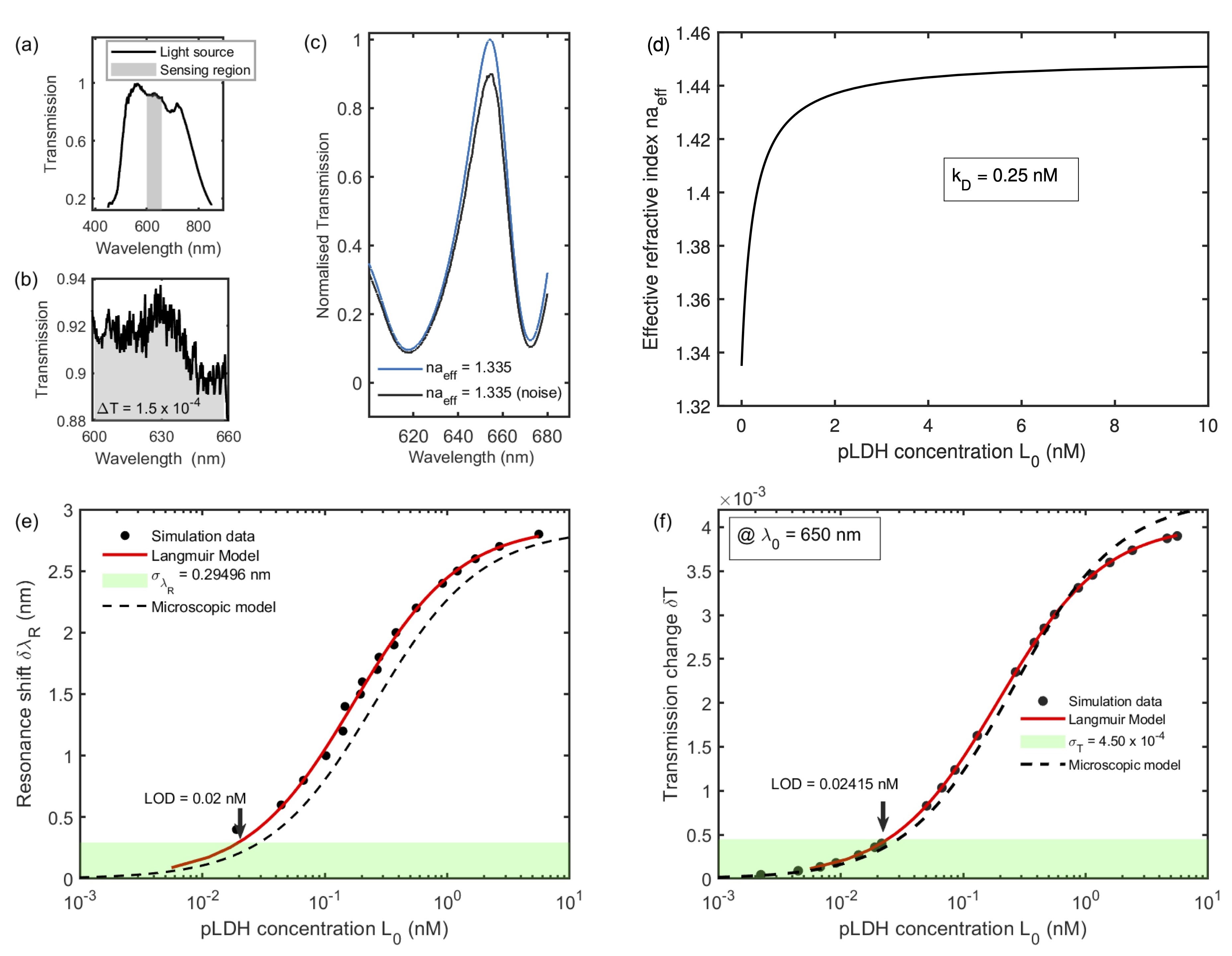}
\caption{\label{model_responce2} Identifying the LOD. (a) Experimental light source spectrum of a white light LED from Thorlabs (MBB1F1). (b) Zoomed-in spectrum, 600 nm to 660 nm, of the light source in panel (a). (c) Normalized simulation spectrum, $na_{\text{eff}} = 1.335$ ($L_0=0$), and light source combined simulation spectrum, $na_{\text{eff}} = 1.335$ (noise). For the noise added spectrum we multiply the normalized `$na_{\text{eff}} = 1.335$' spectrum by the light source spectrum in panel (a). (d) Refractive index $na_{\text{eff}}$ as a function of pLDH concentration, calculated according to the model given in Eq.~\eqref{Maxwell-Garnett_final_new}. (e) A calibration curve to show the resonance shift $\delta_{\lambda_R}$ as $L_0$ changes. (f) The change of transmission $\delta T_{\lambda_0 = 650 \;\text{nm}}$ as $L_0$ increases. The $x$-axis in~(e) and (f) has a log scale. The spectral noise $\sigma_{\lambda_R}$ in panel (e) is calculated from the known $\Delta T = 1.5\times10^{-4}$ according to Eq.~\eqref{delta_lambda}. Further~details about the $\sigma_{\lambda_R}$ and $\sigma_T$ calculation are given in Appendix A, section 2. A Langmuir model, Eq.~\eqref{eq:Lmodel3}, fits the simulation data in (e) and (f). The LOD in both spectral and intensity-based sensing is $\sim 0.02$~nM of pLDH, equivalent to $0.70$~ng/mL. Our analytical microscopic model predicts reasonably well the LOD.}
\end{figure*}

\section{Results}
In Fig.~\ref{model_responce}(a) we show the effect of changing the bulk refractive index $n_d = nb_{\text{soln}}$ of the top dielectric medium of a non-functionalized sensor. Increasing $n_d$ lowers the transmission intensity in the sensing band around peak I ($620-660$~nm). This is because as $n_d$ increases toward $n_s$, the index difference $n_d - n_s$ approaches zero. As a result, the EM-field concentration (bound SP) at the superstrate side is reduced, hence the magnitude of peak I (due to the bound SP at the top nanohole edge) is affected and decreases. To achieve a reasonable transmission within the sensing band, a low index buffer, $nb_{\text{soln}} = 1.335$ (\textit{i.e}. PBS) should therefore be used. The bulk sensing stage is the first stage in most biosensing techniques, where the sensor response is calibrated and its bulk sensitivity measured.

In Fig.~\ref{model_responce}(b), we show sensing of pLDH around peak I for a functionalized sensor. The inset shows a simplified architecture of the functionalized sensor (a part of the nanohole surface). From the plot, it is informative to calculate the local sensitivity, $i.e.$ the change in sensor response $R$ for a given change of the refractive index in the analyte layer, $na_{\text{eff}}$. This local sensitivity,~$S^L_R$, is then given by
\begin{equation}
  S^L_R = \frac{\delta R}{\delta na_{\text{eff}}}=  S_R\frac{\delta n_{\text{eff}}}{\delta na_{\text{eff}}}.
  \label{local sensing}
\end{equation}
For spectral sensing, we have $S^L_{\lambda_R}~=~{\delta \lambda_R }/{\delta na_{\text{eff}}}$. The functionalized metasurface sensor achieves a bulk spectral sensitivity $S_{\lambda_R,\;\text{f}} =236.98$~nm/RIU and a local sensitivity $S^L_{\lambda_R} = 24.58$~nm/RIU, as shown in Fig.~\ref{model_responce}(c) and (d), respectively. For intensity-based sensing, the functionalized metasurface sensor achieves a bulk sensitivity $S_{T,\;\text{f}} =0.4563$~a.u./RIU and a local sensitivity $S^L_{T} = 0.0375$ a.u./RIU, as shown in Fig.~\ref{model_responce}(e) and (f), respectively. Further details about local and bulk sensitivities are given in Appendix A, in Section 1.

In Fig.~\ref{model_responce}(d), we compare the spectral response of the metasurface $\delta\lambda_R$ against $na_{\text{eff}}$ for the simulation and the computed response based on the microscopic model given by Eq.~\eqref{resonance_shift_estimate}. For the microscopic model, input parameters include $nb_{\text{soln}} = 1.335$, $da = 8$~nm, $dc = 15$~nm, $\delta_d = 108.9$~nm, $S_R = 237.09$~nm/RIU, and $na_{\text{eff}}$ was varied from 1.335 to 1.445 to obtain $\delta \lambda_R$. The local sensitivity $S^L_{\lambda_R}$ is the gradient of a $\delta \lambda_R$ versus $na_{\text{eff}}$ graph. $S^L_{\lambda_R} = 24.59$~nm/RIU was obtained for the microscopic model and can be compared to the value calculated based on the simulation of $24.58$~nm/RIU. There is a clear match between the two. Eq.~\eqref{resonance_shift_estimate} is based on planar SPR sensing~\cite{Jung1998}, but we have extended its application to reproduce the sensor response of a nanohole array metasurface. The correct matching of $S^L_R$ is in a sense to be expected since the metasurface supports both bound SPs and propagating SPs, as discussed before. The gap between the two lines (the regression line from the simulation data and the model line) stems from the fundamental difference of the sensor geometries.

The bulk sensitivity, $S_R$, is an intrinsic property of the sensor geometry and optical setup. It is usually measured using a calibration liquid ($e.g.$ water, glucose or glycerol). It is roughly independent of functionalization and is higher numerically than $S^L_R$ (see Eq.~\eqref{local sensing}), but not necessarily more useful in bioassays (see Fig.~\ref{fig1_appendix} in Appendix A). In principle, over a wide range of $\delta n_{\text{eff}}$ and if the sensor geometry and optical setup are unchanged (metal type, thickness, angle, wavelength, etc.), then the functionalized and non-functionalized sensitivities are equal, $S_{R,\;\text{f}} = S_{R,\;\text{nf}}$, as shown in Fig.~\ref{model_responce}(e) for $R=T$. However, they can differ slightly in practice because the functional layer changes the SP field distribution, as shown in Fig.~\ref{model_responce}(c) for $R=\lambda_R$. The parameter $\delta_d$, a key parameter in plasmonic sensing has a local dependency on the analyte adsorbate layer and functional layer~\cite{Beijnum2012}. Adding a thick enough functional layer can significantly reduce the sensor’s ability to detect changes in the surrounding solution. This is often a trade-off for plasmonic sensing: gain specificity (local binding) at the cost of bulk sensitivity, reducing the sensor's sensing ability.

To calculate the LOD, the spectra around sensing peak I are combined with an example light source spectrum to predict an experimental response of the sensor. We modelled a white light source using the measured spectrum of an LED MBB1F1 light source (Thorlabs). In Fig.~\ref{model_responce2}(a), we show the light source spectrum used, and in Fig.~\ref{model_responce2}(b) we show the wavelength range to be combined with the simulation spectra. The normalized spectrum ($na_{\text{eff}} = 1.335$) for a blank solution is shown in Fig.~\ref{model_responce2}(c) as a black line and contains the noise required to calculate the LOD in this study. The transmission noise $\Delta T$ was obtained based on a Matlab algorithm where the moving standard deviation of the broadband light source spectrum was calculated. The average noise is estimated to be $\Delta T = 1.5\times10^{-4}$ a.u. The corresponding spectral noise $\Delta\lambda_R$ is calculated based on Eq.~\eqref{delta_lambda}, where ${\left\vert\frac{dT}{dna_{\text{eff}}}\right\vert_{\lambda_0 = 650 \;\text{nm}}} = 0.0375$~a.u./RIU and $\frac{d\lambda_R}{dna_{\text{eff}}} = 24.58$~nm/RIU are obtained from simulations. Therefore, the spectral noise is $\Delta \lambda_R = 0.09832$~nm. The noise for calculating the LOD is then~$\sigma_{\lambda_R} = 3\Delta \lambda_R = 0.29496$~nm. Using these numbers we can calculate the LOD in terms of $na_{\text{eff}}$. However, it is more informative for clinical purposes to have the LOD in terms of $L_0$.
\begin{table*}[t!]
\caption[]{LOD summary for spectral and intensity sensing comparing simulations with the microscopic model, Eq.~\eqref{resonance_shift_estimate}, for \( f_{\text{max}} = 1 \) and 0.85.}
\label{LOD_summary}
\begin{center}
\begin{ruledtabular}
\begin{tabular}{l *{4}{c} *{4}{c}}
Parameters & \multicolumn{4}{c}{Spectral Sensing} & \multicolumn{4}{c}{Intensity Sensing} \\
\cline{2-5} \cline{6-9}
           & \multicolumn{2}{c}{FEM Simulation} & \multicolumn{2}{c}{Microscopic model} & \multicolumn{2}{c}{FEM Simulation} & \multicolumn{2}{c}{Microscopic model} \\
\cline{2-3} \cline{4-5} \cline{6-7} \cline{8-9}
           &  \( f_{\text{max}} = 1 \)& 0.85          & \( f_{\text{max}} = 1 \) & \(0.85 \) &  \( f_{\text{max}} = 1 \)&0.85         & \( f_{\text{max}} = 1 \) & \(0.85 \) \\
\hline
LOD (nM)   & 0.0198  &0.0238     & 0.0289             & 0.0347             & 0.0242 & 0.0286     & 0.0292             & 0.0351             \\
\end{tabular}
\end{ruledtabular}
\end{center}
\end{table*}

The effective refractive index $na_{\text{eff}}$ is linked to $L_0$ through Eq.~\eqref{Maxwell-Garnett_final_new}, where $f_{\text{max}} = 1$, $K_D = 0.25$~nM, $na_{\text{L}} = 1.45$ and $nb_{\text{soln}} = 1.335$. The corresponding plot for $L_0$ in the range of~$0-10$~nM, is given in Fig.~\ref{model_responce2}(d). A calibration curve to show the sensing performance of the sensor is shown in Fig.~\ref{model_responce2}(e) and (f), where the shift of resonance $\delta\lambda_R$ and the change in transmission $\delta T$ are plotted against $L_0$, respectively. The Langmuir model, Eq.~\eqref{eq:Lmodel3} is then fitted. An LOD~$\sim~0.02$~nM is obtained based on the intersection of $\sigma_{\lambda_R}$ or $\sigma_T$ and the Langmuir fit-line. From Fig.~\ref{model_responce2} (e) and~(f), it can be seen that the analytical microscopic model predicts reasonably well the LOD in both spectral and intensity-based sensing. The results of a further analysis of the LOD based on varying the packing parameter $f_{\rm max}$ are given in Table~\ref{LOD_summary}. Here, the impact on the LOD of sub-optimal packing at the surface of the sensor can be seen as an increase in the LOD value. The trend of the analytical model matches well that of the simulation model when $f_{\rm max}$ is varied. Both models in Table~\ref{LOD_summary} show a $\sim 20\%$ increase in the LOD when using $f_{\rm max} = 0.85$ vs. $f_{\rm max} =1$. 

For the ideal case of $f_{\rm max}=1$, the calculated LOD of $0.02$~nM, equivalent to $0.70$~ng/mL of pLDH ($0.02$~nM $\times$ $35$ kDa = $0.70$~ng/mL), can resolve a submicroscopic malaria infection (parasitemia below $0.001\%$ which is equivalent to $50$ parasites/$\mu$L whose pLDH level is $\sim 1.87$~ng/mL)~\cite{Martin2009, Moody2002}. The average parasite density (parasitemia) seen in patients attending hospitals in
tropical endemic regions is between 5,000 parasites/$\mu$L (0.1\% parasitemia)
and 50 parasites/$\mu$L (0.001\% parasitemia)~\cite{Moody2002}. Using immunochromatographic rapid diagnostic tests (mRDTs) for instance, submicroscopic parasite densities below 100 parasites/$\mu$L are not detected with high confidence due to the threshold of this test. Moreover, patients with parasitemia above 100 parasites/$\mu$L (which is within the mRDT threshold) are sometimes diagnosed negative due to the degradation of the sensor. The LOD for pLDH-based mRDTs is typically in the range of 25~ng/mL~\cite{jimenez2017analytical, jang2013pldh}. Therefore, the theoretical LOD of $0.70$~ng/mL (0.02 nM) we obtain from our model suggests a potential improvement over mRDTs. This would then give a sensor with a more precise and robust indication of early malaria infection. It may also be used to differentiate disease stages based on pLDH levels to initiate proper treatment and intervention.
\begin{figure*}[t]
\centering
\includegraphics[width=0.9\textwidth]{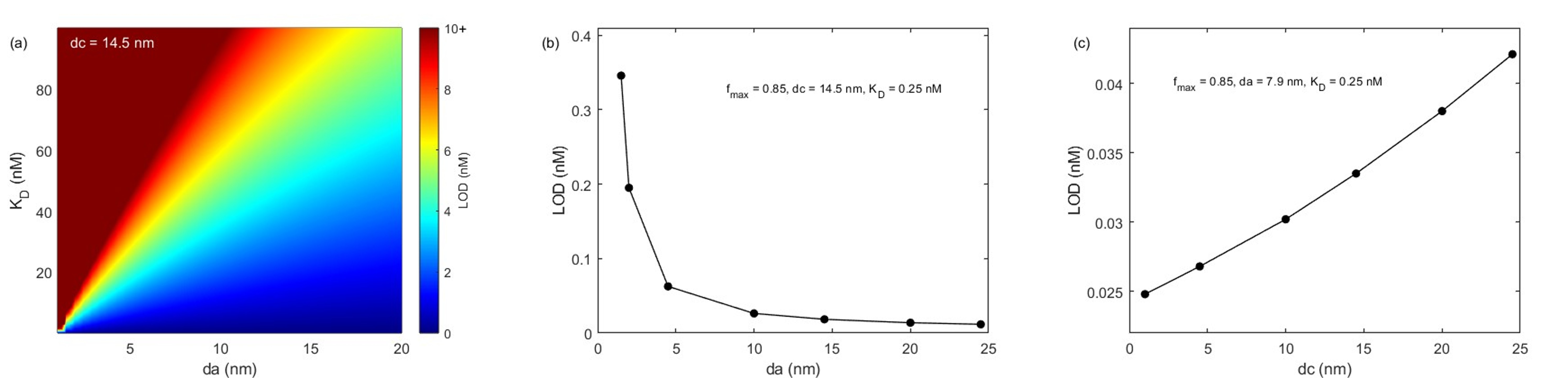}
\caption{\label{practical} Practical consideration of the model for $f_{\text{max}} = 0.85$. (a) LOD color map calculated as a function of dissociation constant $K_D$ and analyte layer thickness $da$. The functionalized layer is fixed as $dc = 14.5$ nm. (b) LOD change with $da$ varied, while $dc = 14.5$ nm and $K_D = 0.25$ nM. (c) LOD change with $dc$ varied, while $da = 7.9$ nm and $K_D = 0.25$ nM.  The model is adaptable by adjusting key parameters like $K_D$, $\;dc$, and $da$, as shown here, to account for functionalization chemistry heterogeneity and analyte orientation.}
\end{figure*}

\section{Practical considerations of the model}

The model's predictions are based on the use of realistic parameters and experimental calibration (or basic FEM simulations), providing a foundation for hypothesis-testing in sensor development. Although the buffer PBS used in our study approximates whole blood lysate in a plasmonic sensing context (as described in Refs.~\cite{lee2012highly, lee2014cationic, vazquez2018enzyme}), real-world factors, such as sensor-to-sensor variability, temperature drift, fluidic instabilities, time span between measurements and non-specific binding may alter the predicted LOD. Our model is adaptable to several of these factors. 

For instance, the parameter $f_{\text{max}}$ can be used to represent the level of heterogeneity of the sensor -- non-uniform binding site coverage due to variations in antibody immobilization or surface chemistry. The results of such a variation are given in Table \ref{LOD_summary}. Furthermore, $\;K_D$ can be adapted for different types of antibodies, and $\; dc$ and $da$ modified for the specific chemistry layer and analyte layer, respectively. The protein refractive index value $na_L=1.45$ taken from Ref.~\cite{barer1954refractometry} may also be an oversimplification and changing this value could improve the model’s predictive accuracy for the LOD through the parameter $na_{eff}$. In addition, in complex matrices like whole blood lysate, biofouling may increase baseline noise~\cite{vasilescu2025promising}. In our model, this can be taken into account to a first approximation by adjusting the buffer solution’s refractive index (\(nb_{\text{soln}} = 1.34–1.36\)). Future work on assessing the accuracy of this modification for model predictability would need to be informed by experimental observations. In Fig.~\ref{practical} we show how the LOD is impacted by some of the parameter variations, where we have set the packing parameter to a more realistic value of $f_{\text{max}} = 0.85$. Parameter adjustments, such as those shown in Fig.~\ref{practical}, may be used to account for sensor-to-sensor variability in order to quantify the model's accuracy in real clinical settings.

Future experiments could validate the microscopic model by functionalizing a NHA (as in Ref~\cite{Cho2013}) with DSP/Protein A/IgG, then spiking PBS with pLDH of different concentration (0.01-10 nM), and measuring spectral shifts via spectrometry. The results can then be compared with the predicted $\delta R$ (from Eq.~\eqref{resonance_shift_estimate}) and used to quantify the accuracy of the model. In this setting, practical LOD values may also increase due to noise from temperature drift ($\sim 0.1$ nm/K~\cite{Jung1998}) and fabrication tolerances ($\pm 10$~nm in hole diameter~\cite{henzie2009nanofabrication}). Further work on including such factors in our model would be needed in this case. Moreover, the single-site Langmuir model used in this work, while foundational for SPR sensor modeling, neglects mass-transport limitations (significant at high flow rates or analyte concentrations), rebinding, surface heterogeneity, and lateral interactions. Future refinements of the model could adopt mass-transport-corrected binding models, two-site Langmuir models for heterogeneous sites, or Frumkin isotherms for intermolecular interactions on the surface. Inclusion of these factors would enhance the accuracy of the model's predictive capability \cite{schuck2010role, urbina2021physical}.

\begin{table*}[t!]
\begin{center}
\begin{ruledtabular}
\centering
\caption{Comparison of our work and other plasmonic enhanced sensors for the detection of malaria.}
\label{tab:comparison}
\begin{tabular}{llccl}
Sensor Type & Target & LOD ({\rm nM}) or ng mL$^{-1}$  & Method & Ref.\\
\hline
\midrule
Au-NHA-SPR & pfLDH/antibody (whole blood lysate) &  Not reported & EOT/Experimental & \cite{Cho2013} \\
Au-NHA-aptasensor & pfLDH/aptamer (in buffer) &  23.5 nM  & Dual-transducer/Experimental & \cite{Bohdan2020} \\
Graphene enhanced SPR & Malaria DNA (in buffer) & 12 pM  & Planar SPR/Experimental & \cite{wu2020ultrasensitive} \\
Fluorescence & pfLDH/aptamer (whole blood) &  18 fM (0.6 pg mL$^{-1}$)  & Plasmon enhanced/Experimental & \cite{Minopoli2020} \\
Au-NHA-SPR & pLDH/antibody (in buffer) & 0.02 nM (0.70 ng mL$^{-1}$
) & EOT/Theoretical Modeling & This work \\
\end{tabular}
\end{ruledtabular}
\end{center}
\end{table*}

It is also interesting to note that metasurface-based sensors can be used to detect multiple analytes within a single test. By building on our results, a multiplexed detection perspective may be explored~\cite{Estevez14,Unser15,Liu20}. Multiplexed detection saves time, lowers cost, and simplifies the diagnostic workflow. Our model can be adapted to multiplexed detection by varying $dc$, $da$, and $K_D$ per subunit of the sensor, each targeting a different biomarker with different surface chemistry. However, multiplexed detection is practically challenging since selective functionalization of subunits (e.g. via microfluidics) may introduce crosstalk. For malaria, multiplexed detection could be accomplished by dividing the metasurface into small subunits and then functionalizing each of them with different antibodies so that the two dominant metabolic proteins (pLDH and pfHRP-2) are detected in parallel. This multiplexed detection would then enable the plasmodium parasite species to be distinguished for effective treatment. Due to non-malarial fevers in the community, rapid and label-free diagnostic biosensors such as the one studied here could also be used in multiplexed form to distinguish and correctly diagnose febrile illnesses from malaria. In areas with high malaria transmission, many febrile illnesses are misdiagnosed as malaria. This can lead to unnecessary antimalarial treatment and delayed treatment for other serious infections. 

Finally, in Table \ref{tab:comparison} we compare the LOD of the Au NHA malaria sensor studied using our model with recent sensors developed and tested for the biological detection of pLDH biomarkers. A more detailed recent summary of the performance of malaria biosensors in general is given in Ref.~\cite{Krampa2020}.

\section{Summary}

Current efforts are being made to develop low-cost optical biosensors based on plasmonic nanotechnology~\cite{shrivastav2021comprehensive, li2019plasmonic}. Here, plasmonic biosensors measure the binding interaction directly as a shift in wavelength or intensity due to a change in the induced surface-bound refractive index. These types of sensors do not use labels and therefore can be applied as an accessible immunoassay nanosensor. For malaria, the most effective strategy largely adopted entails the detection of metabolic proteins and enzymes, most notably pLDH and pfHRP-2~\cite{Ragavan2018, Jain2014}, expressed by the
plasmodium parasite. The protein pLDH is expressed in all metabolically active plasmodium parasite species, whilst an alternative biomarker, pfHRP-2, is only expressed in the plasmodium falciparum species. In this work, we developed and described an effective microscopic model to help build and test a plasmonic sensor capable of detecting pLDH.

More specifically, we characterized a transmission metasurface as a plasmonic biosensor for the detection of pLDH in a buffer solution. The metasurface, modeled using COMSOL, comprised an array of nanoholes patterned in a gold film. We developed an effective microscopic model that can be used to calculate the refractive index of the pLDH adsorbate layer (bound protein layer) from the bulk protein concentration. The effective model is straightforward to use and could be extended to sensing biomarkers of other diseases, such as coronaviruses~\cite{kawasaki2022imprinted} or cancer~\cite{thakur2021vivo,Karki24}, by varying the key parameters $K_D$, $dc$ and $da$. It may therefore help in the future design of functionalized nano-structured SPR sensors without the need for complex numerical simulations. The practical utility of the model is in its ability to aid in the design and testing stages, helping researchers avoid sub-optimal regimes in geometric parameters for a sensor. The predictive advantage of the model is that it provides the achievable LOD for a sensor, giving a tool for carrying out optimisation before complex functionalized experiments are performed. 

For the particular metasurface design we studied, a bulk sensitivity of $S_R=237.09$~nm/RIU, a local sensitivity of $ S^L_R= 24.58$~nm/RIU and an LOD of $0.02$~nM ($0.70$~ng/mL) were obtained. It is important to note that while LODs lower than this, in the fM range, are possible~\cite{Minopoli2020}, these are only achievable using expensive low-light level imaging equipment, putting them out of reach for most labs and clinical settings, as opposed to the more cost-effective metasurface design we have considered. Furthermore, it is important to point out that it is not the low LOD that is of significance in our work, but the theoretical model we developed that can potentially help improve the LOD values of future metasurface-based sensors through careful design, prediction and testing.

The SPR sensor we have modelled, when practically validated using a blood sample, may open up the possibility of using the sensor (and more refined versions of it) in place of mRDTs for the diagnosis of malaria. Moreover, the sensor fabrication and functionalization procedure discussed is simple, effective and does not use labels~\cite{Cho2013}. An extension of the microscopic sensing model to other sensor designs and experimental testing will be critical for future malaria diagnostic applications. 

\acknowledgements

This project was supported by the African Laser Centre (ALC) at the South African Council for Scientific and Industrial Research (CSIR). We also acknowledge support from the University of Dar es Salaam, the Dar es Salaam University College of Education, the Stellenbosch Photonics Institute (SPI) at Stellenbosch University, the Department for Science and Innovation (DSI), and the National Research Foundation (NRF) of the Republic of South Africa.

\appendix
\renewcommand\thefigure{\thesection.\arabic{figure}}  
\setcounter{figure}{0}
\section{}
\subsection{Effective refractive index and plasmonic sensing}
\label{Appendix}
In an SPR sensor, the effective refractive index $n_{\text{eff}}$ is a weighted average of the refractive indices experienced by the SP field. It is given as
\begin{equation}
n_{\text{eff}} = \frac{\int_{0}^{\infty} n(z)|E(z)|^2 \,dz}{\int_{0}^{\infty}|E(z)|^2\,dz},
\label{frist_model1_apped}
\end{equation}
where $n(z)$ is the refractive index as a function of distance from the surface, $|E(z)|^2$ is the intensity of the SP field at depth $z$ and $E(z) = E_0e^{-z/\delta_d}$ is the SP evanescent field which exponentially decays into the dielectric with a decay length $\delta_d$. For visible light the decay length $\delta_d$ ranges between $100 -300$~nm for a planar gold SPR sensor, $5 -50$~nm for a localized SPR sensor (that uses nanoparticles) and $50 - 150$~nm for a nanostructured SPR sensor such as a NHA-SPR-sensor that combines both propagating and localized SPs. Any change in $n(z)$, whether local (near~the~surface) or bulk (throughout) changes $n_{\text{eff}}$, which shifts the sensor response (wavelength or transmission).\\

For a non-functionalized sensor and low concentration $L_0$, $n(z) \simeq nb_{\text{soln}}$, and therefore $n_{\text{eff}}= nb_{\text{soln}}$, where $nb_{\text{soln}}$ is the refractive index of the buffer solution. On the other hand, for a functionalized sensor
$n_{\text{eff}}$ is given by Eq.~\eqref{frist_model11} presented in the main text. The sensor response $R$ is determined by the change in the effective refractive index $n_{\text{eff}}$ and therefore the sensor sensitivity $S_R$ is calculated as 
\begin{equation}
S_R=\frac{\delta R}{\delta n_{\text{eff}}}=\frac{\delta R}{\delta na_{\text{eff}}}\left[\frac{\delta n_{\text{eff}}}{\delta na_{\text{eff}}}\right]^{-1},
\label{Appendix_2}
\end{equation}
where $na_{\text{eff}}$ is the effective refractive index of the bound analyte, calculated using Eq.~\eqref{Maxwell-Garnett_final_new} in the main text, $\frac{\delta R}{\delta na_{\text{eff}}}=S^L_R$ is the `local' sensitivity of the sensor and  $\frac{\delta n_{\text{eff}}}{\delta na_{\text{eff}}} = \Gamma$ is the local sensing factor. For a non-functionalized sensor, $\delta n_{\text{eff}}= \delta na_{\text{eff}} = \delta na_{\text{soln}}$, since analytes are uniformly dispersed in the bulk and therefore $\Gamma = 1$. However, for a functionalized sensor, by using Eq.~\eqref{frist_model11} in the main text, we have
\begin{equation}
\delta n_\text{eff} = \delta na_\text{eff}\left(1-e^{-2da/\delta_d}\right)e^{-2dc/\delta_d}. 
\end{equation}
\begin{figure}[t]
{\includegraphics[width=0.45\textwidth]{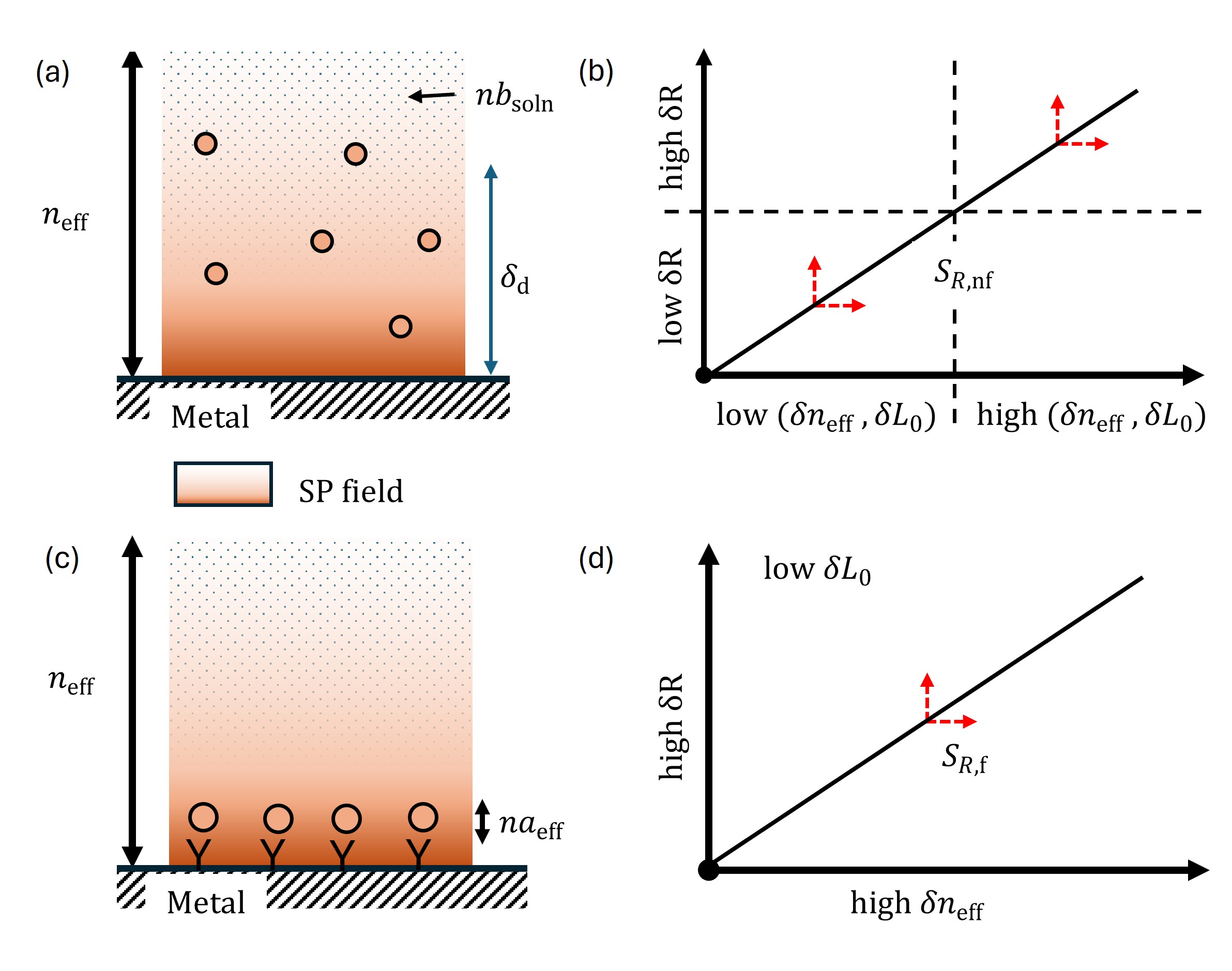}}
\caption{\label{local_sensing} Effective refractive index and plasmonic biosensing. (a) A non-functionalized sensor. (b) Operating regions of a non-functionalized sensor. (c) A functionalized sensor. (d) Operating region of a functionalized sensor. In panels (b) and (d), $S_{R,\;\text{nf}}$ is the bulk sensitivity of a non-functionalized sensor, $S_{R,\;\text{f}}$ is the bulk sensitivity of a functionalized sensor. For the same $\delta n_{\text{eff}}$, $S_{R,\;\text{nf}}=S_{R,\;\text{f}}$. However, at a low concentration change $\delta L_0$, the functionalized sensor offers a higher $\delta R$ due to an enhanced $\delta n_{\text{eff}}$, as a result of analytes being closer to the region with a high SP field, $i.e.$ $na_{\text{eff}} > nb_{\text{soln}}$.}
\label{fig1_appendix}
\end{figure}
Using a Taylor expansion, $1-e^{-2da/\delta_d} \approx \frac{2da}{\delta_d}$, we have $\delta n_{\text{eff}}~\approx~\frac{2da}{\delta_d}\delta na_{\text{eff}}e^{-2dc/\delta_d}$, where $\delta na_{\text{eff}}= na_{\text{eff}}-nb_{\text{soln}}$. Therefore, $\Gamma$ is linear in analyte layer thickness $da$, small for a probe depth $\delta_d \gg da$, and decreases exponentially with thicker functional layer $dc$. This means that large analyte molecules ($e.g.$ proteins) are more easily detected in SPR sensing than small molecule antigens such as steroid hormones \cite{mitchell2010small, englebienne2003surface}. However, by using a localized SP field with $\delta_d \sim 5 - 50$~nm, the local sensitivity is higher than that in a planar SPR sensor.~Also, having a thin functional layer, for instance when using thin cross-linkers ($e.g.$ DSP or 3-GPS) and aptamers instead~of antibodies as receptors, is advantageous. The bulk~and local sensitivities are connected through the relation given in~Eq.~\eqref{Appendix_2} as
\begin{equation}
S_R=S^L_R\left[\frac{\delta n_{\text{eff}}}{\delta na_{\text{eff}}}\right]^{-1},
\label{local_sensor}
\end{equation}
where $S_R$ is defined as a shift in sensor response ($\delta R$) per refractive index unit (RIU) due to change in the effective refractive index $\delta n_{\text{eff}}$ within the whole decay region, and $S^L_R$ is defined as a shift in sensor response ($\delta R$) per change in refractive index $\delta na_{\text{eff}}$ in a thin analyte layer near the surface. The bulk sensitivity, $S_R$, is an intrinsic property of the sensor geometry and optical setup, which is measured or simulated using calibration liquids. It is independent of functionalization and is higher numerically than $S^L_R$ (in Eq.~\eqref{local_sensor}, we have $\left[\frac{\delta n_{\text{eff}}}{\delta na_{\text{eff}}}\right]^{-1}\gg 1$), but not necessarily more useful in bioassays, as shown in Fig.~\ref{fig1_appendix}.

In biosensing, the selectivity and LOD are important factors in addition to the bulk refractive index sensitivity. However, knowing the bulk sensitivity helps predict the local response shift $\delta R$, as given in Eq.~\eqref{resonance_shift_estimate} in the main text, where
\begin{equation}
\delta R = S_R \delta na_{\text{eff}}\Gamma,
\label{gamma_equation}
\end{equation}
with $\Gamma = \left(1- e^{-2da/\delta_d}\right)e^{-2dc/\delta_d}$. Local sensing is more useful because most bio-detection involves surface binding of target analytes ($e.g.$ proteins) in small amounts (weak concentration). With local sensing, the change in $n_{\text{eff}}$ given by $\delta na_{\text{eff}}$, is enhanced even for low analyte concentration, making it possible to detect and quantify analyte binding properties. This is not possible in the non-functionalized case, where a~weak concentration results in a weak $\delta n_{\text{eff}}$ and therefore a weak signal $\delta R$ that is hard to resolve. Let $\delta R_\text{f}$ be the change in response for a functionalized sensor and $\delta R_\text{nf}$ be the change in response of a non-functionalized sensor for same $\delta L_0$. We can use Eq.~\eqref{gamma_equation} for both cases, except for $\delta R_\text{nf}$ we have $\delta na_{\text{eff}}=\delta na_{\text{soln}}$, $dc\to0$ and $da\to\infty$, so $\delta R_{\text{nf}}=S_R\;\delta na_{\text{soln}}$. We then have, for a given change in analyte concentration, $\delta L_0$, $\delta R_\text{f}$ is always greater than $\delta R_\text{nf}$ because of a stronger analyte interaction with the SP field. Taking the ratio of sensor responses, we find
\begin{equation}
\frac{\delta R_\text{f}}{\delta R_\text{nf}} =\frac{\delta na_{\text{eff}}}{\delta na_{\text{soln}}}\Gamma= \frac{na_{\text{eff}}-nb_{\text{soln}}}{\alpha L_0}\Gamma.
\label{Appendix6}
\end{equation}
\begin{figure}[t]
\centering
{\includegraphics[width=0.44\textwidth]{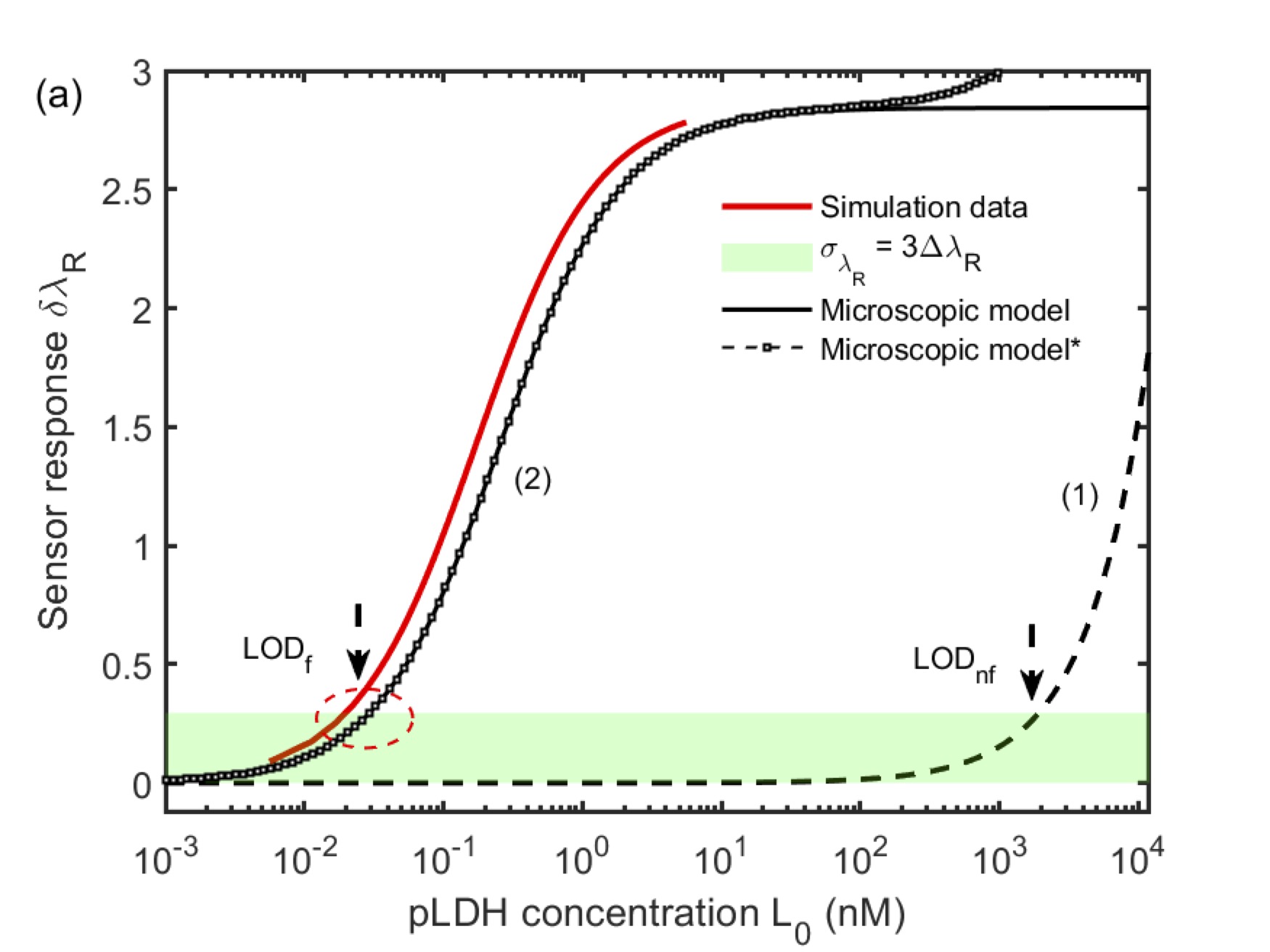}}
{\includegraphics[width=0.44\textwidth]{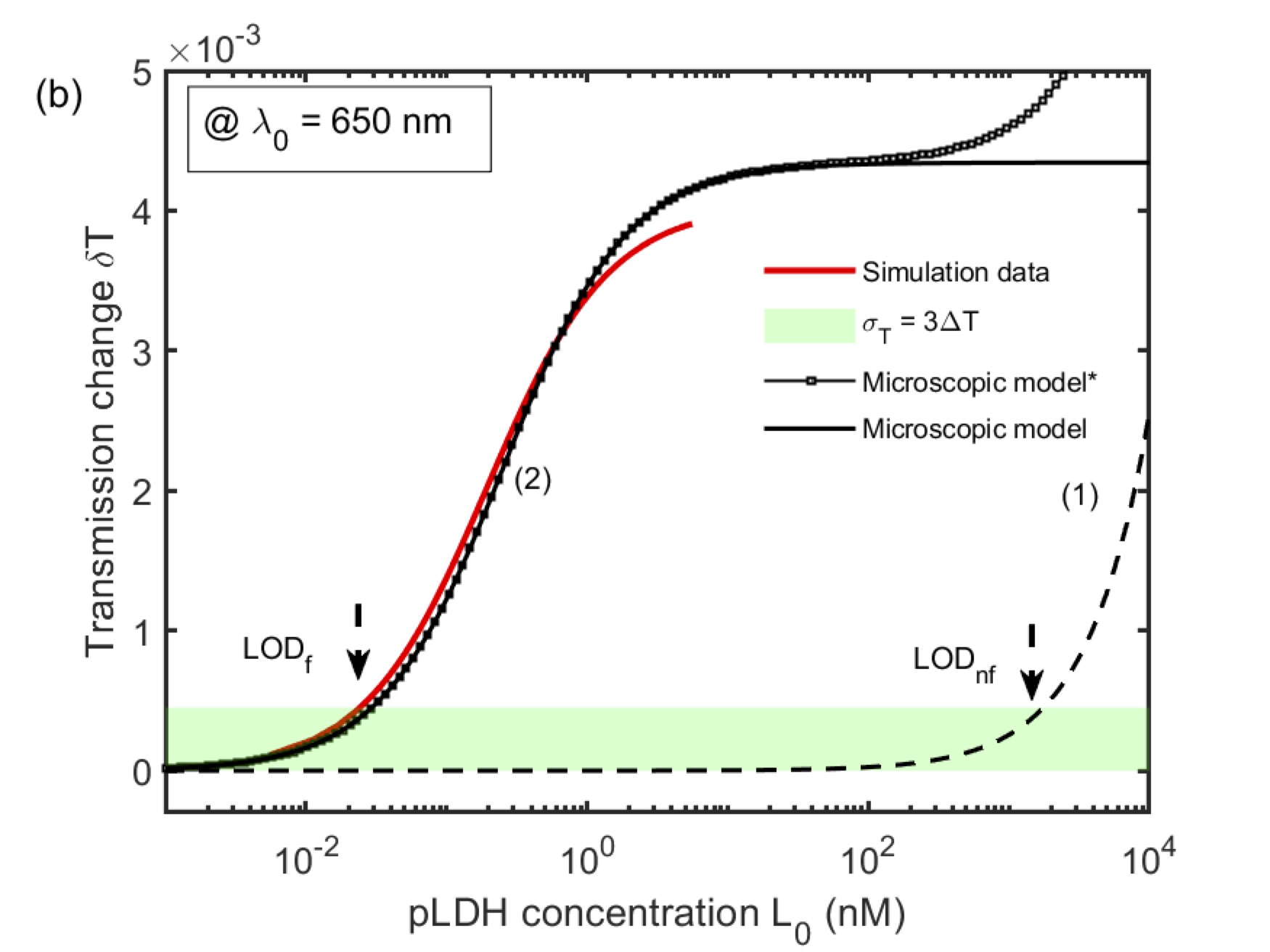}}
\caption{\label{figure3_appendex} Sensor response curve. (a) Spectral sensing. (b) Intensity-based sensing. (1) A non-functionalized sensor. (2) A functionalized sensor. The intersection of the noise line $\sigma_R =3\Delta R$ and the sensor response curve ($\delta R$ versus $L_0$) determines the LOD. At low concentration, and for a fixed sensor noise $\Delta R$, a functionalized sensor offers a better LOD compared to a non-functionalized sensor; $\text{LOD}_{\text{f}}=0.02$~nM vs $\text{LOD}_{\text{nf}}=1900$~nM, respectively. The effective microscopic model using Eq.~\eqref{resonance_shift_estimate} (solid black line) predicts the FEM simulation data and LOD of our metasurface sensor. Microscopic~model* is the effective microscopic model that includes the effect of $\delta na_{\text{soln}}$ for high $L_0$. $\delta na_{\text{soln}}$ has an impact when $L_0$ is high. For high $L_0$, a functionalized sensor behaves like a non-functionalized~one.}
\label{LOD_sketch}
\end{figure}

Consider a pLDH solution, with $L_0 = 250$~nM such that $\alpha L_0 \approx 1.575\times10^{-4}$ RIU, $na_{\text{eff}}=1.45$ and $nb_{\text{soln}} = 1.335$. Using $da = 8$~nm, $dc = 15$~nm, and $\delta_d=108.9$~nm, we have $\Gamma~\approx~0.1037$. Using Eq.~\eqref{Appendix6}, we get $\delta R_\text{f}\approx 75.72\;\delta R_\text{nf}$. With $S_R = 237.09$~nm/RIU and using the relation $\delta R_{\text{nf}}=S_R\delta na_{\text{soln}}$, a 250 nM pLDH solution will result in $\delta_{\lambda_R} = 0.037$~nm if the sensor is not functionalized. On the other hand, if the sensor is functionalized, we find using Eq. \eqref{gamma_equation} the value $\delta_{\lambda_R} = 2.80$~nm. This calculated result for $\delta_{\lambda_R}$ is consistent with our simulation result--shown in Fig.~\ref{figure3_appendex} and Fig.~\ref{model_responce2}(e) in the main text.
\begin{figure}[t]
\centering
{\includegraphics[width=0.43\textwidth]{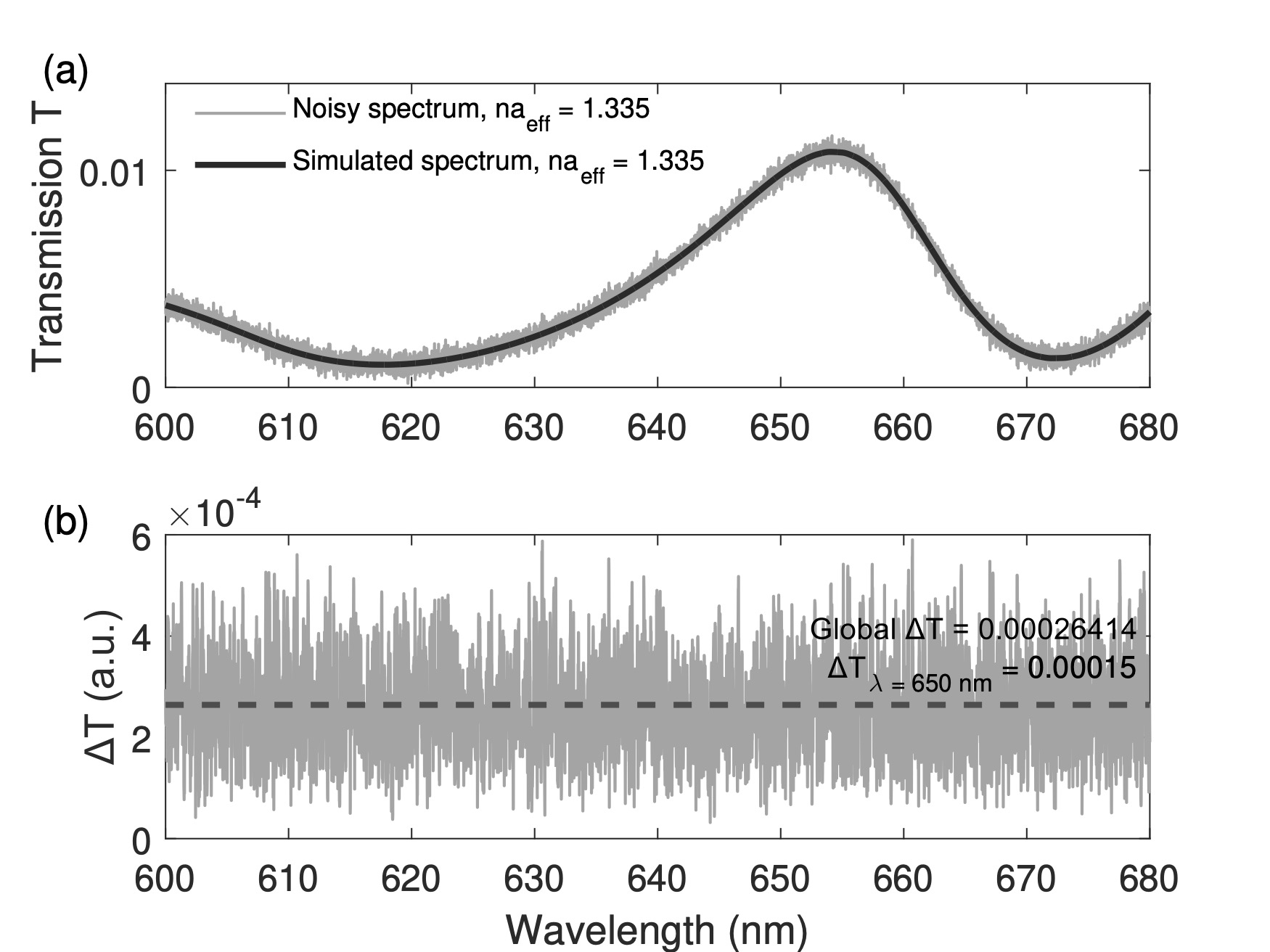}}
\caption{\label{noise_calculation} Global and local standard deviation (STD) in the sensing window $600-680$~nm. (a) The simulated spectrum for $na_{\text{eff}}= 1.335$, $i.e.$ $L_0 = 0$, compared to a noisy spectrum. Noise from the experimental LED spectrum is combined with the simulated spectrum. (b) A 5-point moving STD of the noisy spectrum, $i.e.$ the $\Delta T$ curve. The global STD, $\Delta T = 2.64\times 10^{-4}$ a.u., is the mean STD, and $\Delta T_\lambda =1.5\times10^{-4}$ a.u. at $\lambda =650$~nm is the local STD used to calculate the LOD in the main text.}
\label{fig2_appendix}
\end{figure}

By knowing how functionalization affects the sensor response, we can also show that the LOD calculated for a functionalized sensor exceeds that obtained for a non-functionalized sensor. Let $\Delta L_0 = \sigma_R/S_R$ be the LOD, $i.e.$ the lowest $\delta L_0$ the sensor can resolve in the presence of noise. We know the sensitivity in terms of concentration is $S_\text{$R$,\;nf}=\delta R_\text{nf}/\delta L_0$ for a non-functionalized sensor and $S_\text{$R$,\;f}=\delta R_\text{f}/\delta L_0$ for a functionalized sensor. Since $\delta L_0$ is the same for a non-functionalized and functionalized sensor, we have $S_\text{$R$,\;f}>S_\text{$R$,\;nf}$. For the same $\sigma_R$, we then have $\text{LOD}_{\text{f}}<\text{LOD}_{\text{nf}}$, as can be seen in Fig. \ref{figure3_appendex}. Therefore, biosensing through functionalized sensors provides a better LOD, since the analyte interaction happens in the most sensitive regions with the strongest SP~fields.

\subsection{Noise for LOD calculation}
Here we outline how $\sigma_T$ and $\sigma_{\lambda_R}$ are calculated. We first plot the noisy simulated spectrum in the $600–680$~nm
 range, which is our sensing range. We then calculate the moving standard deviation curve ($\Delta T$) to display both global and local standard deviation of the noisy spectrum. Step-by-step we have: (1) Load data of both the simulated transmission and the experimental LED data and extract data for the $600–680$~nm spectral range. The simulated transmission is shown in Fig.~\ref{fig2_appendix}(a). The LED spectrum (measured at a coarser resolution) was interpolated to match the simulation wavelength grid. (2) Estimate noise from the LED spectrum. The noise level (standard deviation) was estimated using a 5-point moving standard deviation across the interpolated LED spectrum. (3) Generate synthetic noise based on the standard deviation and apply it to the simulated spectrum. The noisy version of the simulated transmission is also shown in Fig.~\ref{fig2_appendix}(a). (4) Compute the moving standard deviation ($\Delta T$ curve). The $\Delta T$ curve was computed using a 5-point moving standard deviation across the noisy spectrum range. (5) Interpolate the $\Delta T$ curve to extract $\Delta T$ at $650$~nm. This is shown in Fig. \ref{fig2_appendix}(b). The global $\Delta T$ was calculated as $\Delta T = 2.64\times 10^{-4}$ and the local value as $\Delta T_{\lambda = 650 \;\text{nm}} = 1.5\times10^{-4}$. For the LOD calculation, $\Delta T_{\lambda = 650 \;\text{nm}}$ was used and the corresponding noise terms~based on the formula $\sigma_R = 3\Delta R$ are $\sigma_T = 4.5\times10^{-4}$ and $\sigma_{\lambda_R} =0.29496$~nm, for intensity and spectral sensing,~respectively.

\bibliography{Final_version_v2}

\end{document}